*CTransformer*: Deep-transformer-based 3D cell membrane tracking with subcellular-resolved molecular quantification

Zelin Li[1]‡, Guoye Guan[2,3]‡§, Xiu Xian[1]‡, Dongying Xie[4], Yiming Ma[4], Sicheng You[1], Zhen Zhu[1], Darrick Lee[5], Zirui Zhang[1], Zhuohen Ran[1], Chenwei Wang[1], Jianfeng Cao[6], Chao Tang[7,8,9,10], Zhaoke Huang[1]§, Zhongying Zhao[4,11]§, Hong Yan[1]§

1. *Department of Electrical Engineering, City University of Hong Kong, Hong Kong SAR, China*
2. *Department of Systems Biology, Harvard Medical School, Boston, USA*
3. *Department of Data Science, Dana-Farber Cancer Institute, Boston, USA*
4. *Department of Biology, Hong Kong Baptist University, Hong Kong SAR, China*
5. *Mathematical Institute, University of Oxford, OX2 6GG, United Kingdom*
6. *School of Biomedical Engineering, Harbin Institute of Technology (Shenzhen), Shenzhen, China*
7. *Peking-Tsinghua Center for Life Sciences, Peking University, Beijing, China*
8. *School of Physics, Peking University, Beijing China*
9. *Center for Quantitative Biology, Peking University, Beijing, China*
10. *Center for Interdisciplinary Studies, Westlake University, Hangzhou, China*
11. *State Key Laboratory of Environmental and Biological Analysis, Hong Kong Baptist University, Hong Kong, SAR China*

‡   *Equally contributions*
§   *Corresponding authors:*

guanguoye@gmail.com (G.G.); zhahuang@cityu.edu.hk (Z.H.); zyzhao@hkbu.edu.hk (Z.Z.); h.yan@cityu.edu.hk (H.Y.)

## Abstract

Deep learning segmentation and fluorescence imaging techniques allow the cellular morphology of living embryos to be constructed spatiotemporally. These development processes involve numerous molecules distributed at the subcellular scale, such as cell adhesion (E-cadherin), which accumulate at cell-cell interfaces to regulate intercellular connection. However, quantifying molecular distributions within specific subcellular regions across the entire embryo — where cell movement and molecular redistribution occur rapidly — is challenging due to the need for simultaneous cell morphology reconstruction and lineage tracing due to photobleaching and phototoxicity. We report a transformer-based pipeline, *CTransformer*, that establishes a 4D cellular morphology map before the 550-cell (late) stage. *CTransformer* constructed 4D cellular morphology atlases, reaching 80% accuracy at the 550-cell stage. Through this advanced architecture, we use only one channel to reconstruct cell morphology and achieve cell tracing. With each cell's morphology as a reference, the distribution of specific molecules throughout the cell body and at cell interfaces can be quantitatively measured in another fluorescence channel. We apply this methodology to track E-cadherin during embryonic development of the worm *Caenorhabditis elegans*, from fertilization to gastrulation. Our results reveal that E-cadherin is tightly regulated across individual embryos, both within single cells and at cell-cell interfaces, displaying an anterior-posterior gradient and cell- and lineage-specific patterns. Furthermore, its spatiotemporal heterogeneity influences cell mechanics and embryonic morphogenesis, helping explain how *C. elegans* achieves stereotypical developmental patterns at cellular resolution. In summary, *CTransformer* offers a high-throughput approach for measuring molecular distributions at subcellular resolution, providing valuable knowledge for cellular and developmental biology, biophysics, and bioinformatics.

**Keywords:** Cell Membrane, Cell Morphology, Molecular Quantification, Subcellular Resolution, E-cadherin, Cell Adhesion, Transformer, Generative Adversarial Network (GAN)

# MAIN TEXT

## Introduction

The spatial dynamics of molecules impact a lot of physiological functions from the subcellular to organismic scale. For instance, in the mechanical sense, the life-dependent periodic exchange of actin filament (F-actin) between the cytosol and cortex of a cell determines its stiffness around cell division [1]. The epithelial cadherin (E-cadherin) located on the cell membrane is affecting not only the contact between neighboring cells but also drives collective cell sorting. In the chemical sense, the phase separation and movement of specific molecules inside a cell, such as the P granule that maintain the stemness in germline cells, is critical for specifying the stemness in its daughter[2]. Such polarized behavior is guided by the upstream reaction-diffusion molecule network that sets up gradients and cell polarization [3]. Moreover, the many important subcellular structures are in drastic spatial transformation that correlates to the physiological state of a cell, such as the centrosomes that drive cytokinesis. This numerous information at the subcellular scale is hard to monitor especially in a crowded multicellular system like a rapidly developing embryo, where the cell membrane morphology is usually hard to recognize and track simultaneously.

Cell morphology (shape, surface and contacts) cannot be obtained simultaneously with other markers because of the number of channels needed and the problems of photobleaching and phototoxicity [,12]. So specific expression of some properties, *e.g.*, membrane adhesion, can only be analyzed by approximate locations of nuclei rather than by precise cell morphology. If cell shapes and lineages can be traced with only one protein marker in *in vivo* embryos, photobleaching and phototoxicity will be lowered, keeping cells healthier and closer to "wild-type" behaviour. An extra channel can then observe other proteins, such as globular multi-functional protein family members, e.g. Actin, or cell adhesion proteins, e.g. *hmr-1*, a kind of *E-cadherin*, to coordinate developmental processes with the cell morphology map. Thus, conducting cell shape segmentation and cell identity tracing simultaneously via the 3D cell membrane tracking (for cell morphology map) is desired. This is critical for reconstructing the spatial domain of cells and supports the quantification of molecules at subcellular scale. Although with tuned imaging balances, the advanced optical and deep learning segmentation algorithms are established [4-7], recognizing cell membrane signals and reconstructing morphology maps are far from perfection. Under normal circumstances, the maps are derived from membrane and nucleus fluorescently labeled images, giving the cell positions and detailed cellular shape. However, the two missions are usually realized by two complementary fluorescence channels: while the cell nuclei with condensed fluorescence intensity serve as the seed for cell identity tracing, they also guide cell membrane segmentation and assign cell identity for each cell region [8]. This significantly limits the real-time monitoring of molecules of interest in the additional fluorescence channel. Therefore, realizing these two tasks to track cell membranes is a technical solution not only making full use of image information but also providing one more fluorescence channel for monitoring a molecule. Recent works were performed on ascidian embryogenesis but still rely on manual correction.

For discovering shape-expression patterns, membrane-labelled fluorescent signals are indispensable to build morphology maps. Although a widely applicable and label-free method for multiple organisms is used for tracing in early cell stage[9], a more accurate nuclei-marker-independent method with short manually correction time is missing, especially at late-cell stage (to and over 550 cells). By building a morphology map with nuclei-independent images, it should be possible to build an expression map along a single-cell shape. In previous studies, other gene-marks or proteins are estimated along the cell nuclei map only[10, 11]. Cell morphology (shape, surface and contact) cannot be derived simultaneously with other markers because of the number of channels, photobleaching and phototoxicity[6, 12]. So previously specific expressions are analyzed only along with approximate nuclei locations rather than on precise cell morphology[10, 13]. By generating both the cell shape and the lineage tracing with only one protein for *in vivo* embryos (no nuclei fluorescence staining), first we can lower the photobleaching and phototoxicity effects to keep healthier and more "wild-type" embryos; second, we can use the extra channel to observe other proteins based on accurate cell shape, like globular multi-functional protein family, e.g. Actin, or cell adhesion protein, e.g. *hmr-1*, a kind of *E-cadherin* (see Figure 1, section Single-cell Adhesion and Cell-cell Contact Adhesion), to observe some developmental processes with cell morphology map. Now, with the proposed *CTransformer*, the actual and real expressions on the segmented shape, cell surface and cell-cell interaction contact in *in vivo* embryos can be quantitatively calculated and analyzed. This allows researchers to carry out previously unachievable experiments to discover regulations between cell shape and other expressions. Like cell adhesion[14, 15], a fundamental property in multicellular development, whose surface-wise and contact-wise patterns in the living structural systems can/are not studied before.

In downstream biological analyses, incorrect cell shapes could trigger a chain of errors, which requires most accurate cell morphology segmentation[6, 12, 16-19]. With current adaptive scanning fluorescence microscopes (advanced hardware modalities) [20, 21], efforts to improve cellular recognition and segmentation focus on enhancing image clarity (super-resolution), resolving ambiguous signals, and generating instances (cell membrane recognition and instance segmentation). However, with nuclei staining, inherent fluorescence heterogeneity and weak membrane signals in edge cells remain challenging. The trade-offs between phototoxicity and image quality lead to laser attenuation, affecting imaging in ambiguous regions and introducing anisotropy, particularly along the axial direction, which impacts recognition and segmentation over time. Additionally, late-stage embryos contain numerous cells with irregular shapes and anisotropic sizes, making it challenging to accurately map single-cell morphology in dense cell structures. Even minor errors in cell membrane recognition can result in large areas of incorrectly connected or missing cells, and such errors may be exponentially amplified. While the cell shape segmentation using only cell membrane information is challenging, following cell identity tracing on top of it further requests its high accuracy as well as a robust tracing algorithm based on 3D cell regions, instead of the usual one on top of cell nuclei.

Membrane-labelled fluorescent signals are required to build morphological maps to discover shape-expression patterns, but can cellular morphology and lineages be determined with only membrane-marked images, as this could maintain cell health and potentially give complete embryonic data? A widely applicable and label-free method suitable for multiple organisms has been used for tracing early embryonic cell stages[9]. More accurate nuclei-marker-independent methods with short manual correction times, especially for late-cell stage embryos (to and over 550 cells) are still needed. Building a morphological map with nuclei-independent images should allow a genetic expression (measuring cellular properties) map of single-cell shapes to be built. In previous studies, expression of gene-markers or proteins were estimated only following the map of cell nuclei map[10, 11].

In this paper, we introduce a generalizable and adaptive integrated DNN pipeline *CTransformer* for 4D image cellular segmentation, nuclei-independent tracing (membrane lineaging) and shape-expression analyses. With *CTransformer*, the actual expression of specific genes and proteins associated with segmented shape, cell surfaces and cell-cell interaction contacts in *in vivo* embryos can be quantitatively calculated and analyzed. This allows previously unachievable experiments to discover regulatory patterns between cell shape and other cell membrane features like cell adhesion[14, 15], a fundamental property in the development of multicellular organisms where surface shape and contact patterns are informative. One channel is used to segment every single-cell's shape through fluorescent membrane signals. Centroids of the shapes are used to generate images of pseudo nuclei for efficient lineage tracing. Another channel measures the intensity of each cell's adhesions, allowing discovery of patterns (Figure 1). The DNN model of *CTransformer* for segmentation by morphology is a semi-supervised Topology-constraint U-shape Nucleus-prompting Euclidean distance transform Transformer (*TUNETr*). To ensure that *TUNETr* performs well with limited GT training, *TUNETr* integrates the relative positional attention mechanism of transformers[29], prior inductive bias knowledge and image processing technologies in a U-shaped network, topology-constraint loss function[30, 31]. Boundary-aware semi-supervised learning and synthetic GT data are utilized to solve the challenges in failure to segment axial and unseen images. These modules together form a human-in-the-loop pipeline to automate cell segmentation based on morphology. The potential of *CTransformer* for 4D *in vivo* cellular segmentation with *C. elegans* (Movies S1 and S2), is shown by time-series of 48 embryos. Moreover, A membrane to nucleus Generative Adversarial Network (*m2nGAN*), a generative model to artificially generate images of pseudo nuclei from cell shapes is developed to allow tracing of cell lineages from only membrane images. *CTransformer* was applied to the segmentation of *C. elegans* live embryos, previously imaged and annotated by 5 biological experts, under 3 different imaging conditions, with 6279 and 2339 cell objects, for training and evaluation respectively. Segmentation of the *in vivo* embryos was evaluated by multiple metrics, in comparison with 9 existing DNN-based methods, on 30509 cell regions, which demonstrated its superior performance. The images of pseudo nuclei for *C. elegans* were tested and showed effectiveness in tracing cell lineages up to 550-cell stage embryos. A user-friendly interface and feasible scripts are provided for users, allowing researchers to run segmentation and visualization of the morphology map in ITK-SNAP-CVE[12]. The proposed pipeline could pave the way for tracing of cell lineages using membrane images, and enable discoveries of more transcriptomic, e.g. adhesion expression, shape-related patterns.

4D fluorescence images, also described as 3D + *t* data

# Result

***CTransformer* pipeline is designed to track 3D cell membrane fluorescence images, enabling both accurate cell morphology segmentation and cell lineage tracing.**

*CTransformer* is a fully automated and integrative experimental-computational framework designed to track time-lapse 3D cell membranes and reconstruct both cell morphologies and lineages across entire samples, using only the experimental fluorescence channel of cell membranes, which quantifies spatiotemporal dynamics of molecules of interest on top of the reconstructed cell morphologies and lineages simultaneously using another fluorescence channel (Figure 1). In the first channel, 3D + *t* images of cell membrane fluorescence are subjected to morphology segmentation (*TUNETr* Module - a <u>d</u>eep <u>n</u>eural <u>net</u>work or DNN for voxel-wise segmentation of cell membranes) (Figure 1a); then the segmented morphologies are processed to generate cell nuclei pseudo-fluorescence subjected to lineage tracing, without reliance on the experimental fluorescence channel of cell nuclei as before (*m2nGAN* Module - a generative <u>a</u>dversarial <u>n</u>etwork or GAN translating membrane-segmented images into nuclei-mimicked images) (Figure 1b). Building upon the reconstructed whole-sample cell morphologies and lineages, the additional channel(s) of 3D + $t$ images of labeled molecule(s) provide its quantitative distributions of molecular signals at any spatial and time points (*MolQuantifier* Module - an adjustable combination of molecular measurements specialized for spatial regions, including but not limited to those accumulated on specific cell surfaces and cell-cell contact interfaces). Altogether, the full utilization of multiple fluorescence channels enables *CTransformer* to automatically delineate the complicated spatiotemporal dynamics of entire biological samples, spanning from molecular, subcellular, cellular, to multicellular scales.

To satisfy the segmentation accuracy requirement for *CTransformer* construction, we firstly established a well-organized dataset for multiple *in vivo C. elegans* embryos documented with unified data format (see Materials and Methods). This dataset contains the 3D + $t$ images (a total of 16,922 dual-channel volumetric images) of 34 compressed embryos (*incl.*, 29 wild-type samples developing normally and 5 RNAi-treated samples developing abnormally) and 14 uncompressed (*incl.*, 9 wild-type sample developing normally and 5 RNAi-treated samples developing abnormally) embryos for establishing the *TUNETr* Module and *m2nGAN* Module. Those embryo samples are fluorescently labeled by green <u>f</u>luorescent <u>p</u>rotein or GFP to label cell nuclei (stimulated by laser beam at 488 nm) and mCherry to label cell membranes (stimulated by laser beam at 594 nm), covering *C. elegans* embryogenesis from 2- to >550-cell stage, namely, from the completion of the first cell division to the last round of most cell divisions. While a part of embryo samples are adopted from previous public dataset for enlarging training data pool and is subject to new segmentation/tracing with higher quality or yet to be done[6, 12], another part are newly published in this paper to test *CTransformer*'s general applicability (Table S1). Last but not least, a new strain were newly crossed with mCherry labeling cell membranes and GFP labeling adhesive protein HMR-1/E-cadherin , producing three additional embryos are were imaged from the fertilization to gastrulation, for establishing the *MolQuantifier* Module (see Materials and Methods).

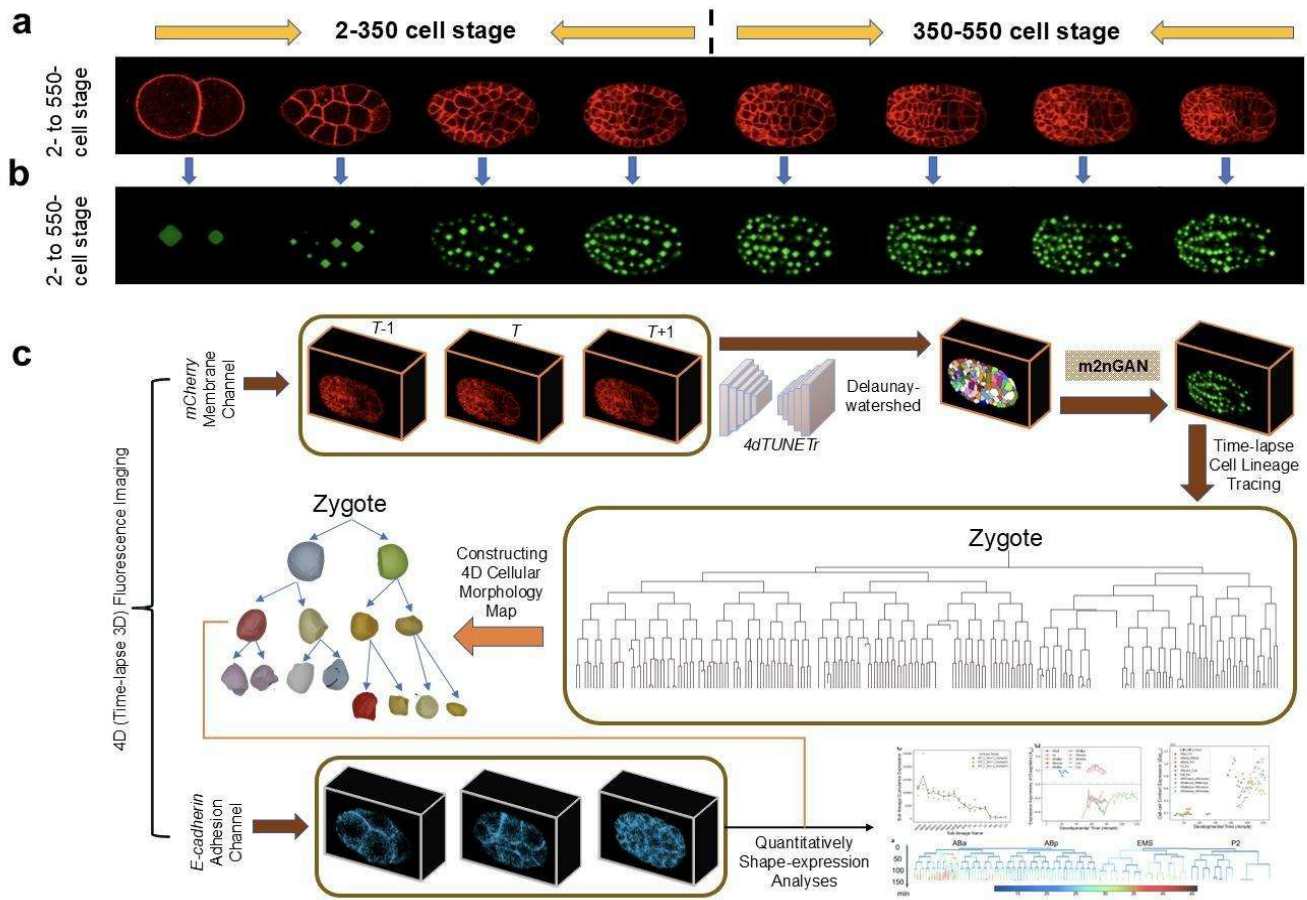

**Figure 1. Experimental-computational design of the integrative *CTransformer* framework, exemplified by its application to worm *C. elegans* embryogenesis.** (**a**) *in vivo* collection of 3D + *t* images about fluorescence labeling cell membranes (displayed with 2- to 550-cell stages at top and >550-cell stage at bottom, when the fluorescence signal becomes progressively ambiguous and difficult to recognize. (**b**) *in silico* generation of 3D + *t* images about pseudo-fluorescence mimicking cell nuclei. (**c**) *CTransformer* workflow from cell membrane fluorescence images to cell nuclei pseudo-fluorescence images, which are subsequently used for cell morphology segmentation (*TUNETr* Module) and cell lineage tracing (*m2nGAN* Module), respectively, ultimately enabling whole-sample, subcellular-resolution quantification of molecules of interest (exemplified by the adhesive protein HMR-1/E-Cadherin) across space and time (*MolQuantifier* Module).

***CTransformer* segments cell morphologies accurately, consistently performing well even under conditions of large cell number and high cell density.**

**The first module of *CTransformer* is TUNETr:** a scalable and high-capacity vision model, inspired by SwinUNETR, specifically tailored for voxel-wise segmentation in large-scale embryonic imaging datasets. It is designed to capture complex 3D morphological features while maintaining robustness across a wide range of developmental stages and imaging conditions. The model is initialized with *iTUNETr*, a backbone pre-trained on a curated set of manually annotated 3D volumes and progressively refined through two specialized variants: *sTUNETr* (for static 3D snapshots) and *4dTUNETr* (for continuous time-lapse data). These variants are trained using a semi-supervised learning framework that incorporates edge-aware supervision, pseudo-label propagation, and topology-informed structural priors (Figures. S1, S2a). This multi-stage training strategy, anchored by expert-curated cell lineage trees, enables precise boundary delineation and ensures robust generalization across diverse spatiotemporal imaging conditions.

The potential of *CTransformer* for accurate cell segmentation was tested on the dataset (named "*3D Shape Evaluation*" dataset accessible via https://doi.org/10.6084/m9.figshare.27085657), which includes manually annotated ground truth (GT) from 16 distinct 3D volumetric images with 2,339 3D cell objects. These annotations were curated to represent significant developmental milestones, providing a robust basis for evaluating segmentation capabilities prior to the 550-cell stage. *CTransformer* was compared with a series of state-of-the-art algorithms *3DUNet++*, *VNet*, *Swin UNETR*, *Cellpose3D*, *StarDist3D*, *CShaper*, and *CShaper++*. Each method was tested under identical conditions, ensuring a fair and unbiased comparison; their performance was evaluated by the Dice Score, Hausdorff Distance, and Jaccard Index, and compared with each other.

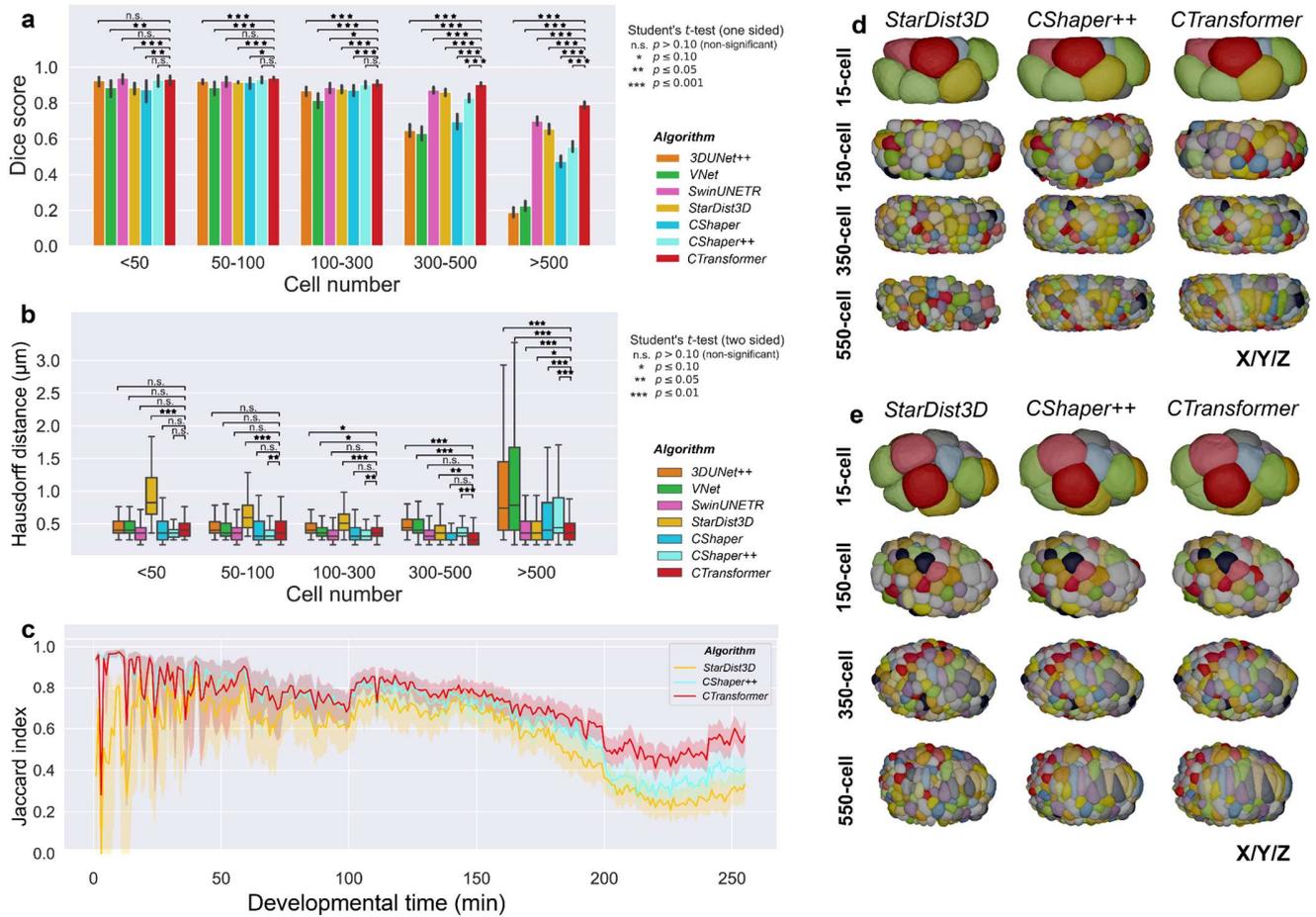

**Fig. 2 | Outperformance of *CTransformer* compared with other state-of-the-art algorithms, in terms of segmentation quality. a**, Dice score over developmental stage defined by cell numbers, demonstrating *CTransformer*'s outperformance at late developmental stages. (**b**) Hausdorff distance at different developmental stages defined by cell numbers, demonstrating *CTransformer*'s outperformance at late developmental stages. (**c**) Jaccard index over developmental stages defined by time point, demonstrating *CTransformer*'s outperformance at late developmental stages. (**d**) Segmentation results of a compressed embryo at its 15-, 150-, 350-, and 550-cell stages (illustrated with the embryo sample WT_C_Sample2). (**e**) Segmentation results of a compressed embryo at its 15-, 150-, 350-, and 550-cell stages (illustrated with the embryo sample WT_Sample1).

Dice Score is a statistical and pixel (voxel) based measure that assesses the similarity between predicted cells ($Cell_{pred}$) and ground truth cells ($Cell_{gt}$). It is defined as twice the area of overlap between the two segmentations divided by the total number of voxels in $Cell_{pred}$ and $Cell_{gt}$:

$$Dice = \frac{2 \times |Cell_{pred} \cap Cell_{gt}|}{|Cell_{pred}| + |Cell_{gt}|} \tag{1}$$

*CTransformer* achieved an overall value of 0.908 ± 0.003, the greatest accuracy in embryonic cell segmentation and 0.02 higher than the second-best method, *SwinUNETR*, which achieved a Dice score of 0.888 ± 0.006 (Figure 2a). This metric, indicative of the overlap between predicted and actual cell segmentations, highlights

*CTransformer*'s effectiveness in capturing cell morphology (Table S2). (Table S3). While *CTransformer have comparable performance with 3DUNet++, CShaper++, SwinUNETR, it* significantly outperforms all other algorithms for from 300- to beyond 550-cell stage.

**Hausdorff Distance** measures the greatest distance from all voxels in the surface of $Cell_{pred}$ to their closest voxel in the surface of $Cell_{gt}$. It is calculated as:

$$H(Cell_{pred}, Cell_{gt}) = max\left(h(Cell_{pred}, Cell_{gt}), h(Cell_{gt}, Cell_{pred})\right) \quad (2)$$

where $h(Cell_{pred}, Cell_{gt}) = |a - b|$ and $h(Cell_{gt}, Cell_{pred})| b - a|$.

*CTransformer* had a score of 2.591 ± 0.187 µm (Table S4), indicating minimal boundary discrepancies and higher fidelity in representing cell shapes. This performance was generally better than other methods, with SwinUNETR, the second best, scoring 2.598 ± 0.339 µm (Figure 2b).

The **Jaccard Index** (JI), also known as Intersection over Union (IoU), was computed for successfully segmented cells and averaged across all cells. It is demonstrated as:

$$Jaccard\ Index = \frac{True\ Positive\ (TP)}{TP + False\ Positive\ (FP) + False\ Negative\ (FN)}$$

These metrics were selected for their ability to thoroughly evaluate performance of segmentation across temporal sequences and varying cell stages, reflecting the complexity of embryonic development. all state-of-the-art methods experienced a decline in accuracy over time, (~359 minutes from the 4-cell stage to the 589-cell stage). However, *CTransformer*'s degradation was the smallest, maintaining the highest accuracy across different developmental stages. It achieved a Jaccard Index of 0.715 ± 0.02 (Table S5), outperforming StarDist3D (0.563 ± 0.033; Table S6) and *CShaper+* (0.673 ± 0.037; Table S7).

The time-lapse curves of two embryos further demonstrated *CTransformer*'s temporal robustness and consistency throughout embryonic development (Figure 2, d and e).

All the above three metrics coherently highlight CTransformer's unique segmentation performance especially for **conditions of large cell number and high cell density.**

Overall, *CTransformer* outperformed other deep learning-based methods by over 10% for construction of cell morphology maps, as shown at five key developmental stages. These results were consistent across various imaging conditions (two confocal systems), culture situations (compressed or uncompressed), and different scanning depths (68, 70, and 92 layers).

In addition to single time-point evaluations, time-lapse assessments of embryonic development were assessed to evaluate overcoming challenges associated with constructing comprehensive 2D + *t* ground truths. The **Time-lapse 2D Evaluation** dataset (https://doi.org/10.6084/m9.figshare.2708565) consisted of two embryos with a total of 460 images and 30,509 cell regions. Annotating the central 2D YZ plane cut from over 200 volumetric series, gave a robust way to evaluate segmentation performance over time. *CTransformer* demonstrated strong temporal robustness, as evidenced by its superior Jaccard Index compared to established methods like StarDist3D and *CShaper++*.

Qualitative comparisons between StarDist3D, *CShaper++*, and *CTransformer* are presented in Figure 2 (d and e) and Figure S3. Regardless of compression conditions, cell stages (350- or 550-cell), or imaging planes (axial or sagittal), *CTransformer* consistently made fewer visual errors.

With above results, in summary, the *CTransformer*'s **TUNETr** module is an effective tool for segmentation of live-cell embryos by 4D morphological mapping. Its proficiency across various metrics and conditions represents a significant advance for the accurate portrayal and understanding of cellular dynamics. *CTransformer*'s application in biological studies will open avenues for future exploration into its potential enhancement and broader implications for 3D biological imaging.

## With only membrane FC images, *CTransformer* tracks both cell morphology and lineage across multiple dimensions throughout embryonic development with high fidelity.

In this section, achieving low rates of cell loss during the fluorescent imaging and optical scanning of 3D embryos is described. 4D images may encounter intrinsic light heterogeneity and weak signals, as in the inner and middle cell layers due to larger inner cells and less dense expression of signaling proteins (*mCherry*) (Figure 4a). This anisotropy can lead to failures in segmentation and cell loss. relative to other methods (Figure 3a, Figure S6), *CTransformer* showed lower anisotropy in brightness, more balanced signal strength across different cell layers, and better recognition of 3D membrane at attenuated laser intensities. *CTransformer* effectively resolved weak boundaries (at the exterior of the embryo) by focusing on thickening outer membranes (Figure S6). The topological loss helped connect incomplete and invisible membranes by learning the embryo's membrane skeleton in a topological latent space. Here, for this intermediate step in recognition of membranes, the reconstruction of 3D cell membranes in ambiguous images was improved for consecutive single-cell segmentations, especially or late-stage embryos with small cells.

The performance of *CTransformer* surpassed that of other methods in reconstructing single-cell morphology with minimal cell loss across time, tissue type, and lineage (Figure 3, c to e). The evaluation was conducted on 2 wild-type compressed and 1 uncompressed *C. elegans* embryo. The rate of cell loss was determined by comparison against manually traced nuclei numbers to assess the proportion of missing cells in predictions. *CTransformer* successfully generated 96.5% of cell objects, while StarDist3D and *CShaper++* constructed 62.0% and 83.7% of cells in the two evaluated embryos (Table S8).

*CTransformer* could reconstruct skin cells to the 550-cell stage without errors but there were some errors after that[38] (Figure 3c), with low loss rates (around 1%), achieving a tenfold improvement over other methods. The average loss rate of cells of extremely small size (less than $10 \mu m^3$), like apoptotic cells, was reduced to ~20% (Figure 3d). Different cells could be better constructed in the morphological map by *Ctransformer* enabling more feasible studies of biological processes, like apoptosis, with support from the more accurate detailed cell-cell contact atlas. In lineage assignment, *CTransformer* dealt with deformed and small sizes with lower loss rates, less than 15% for progeny of the AB and C lineages (Figure 4e, Table S8.

Cells segmented by *CTransformer* exhibited better consistency in recognition of the cells and their shapes over time in the real embryonic cell map.

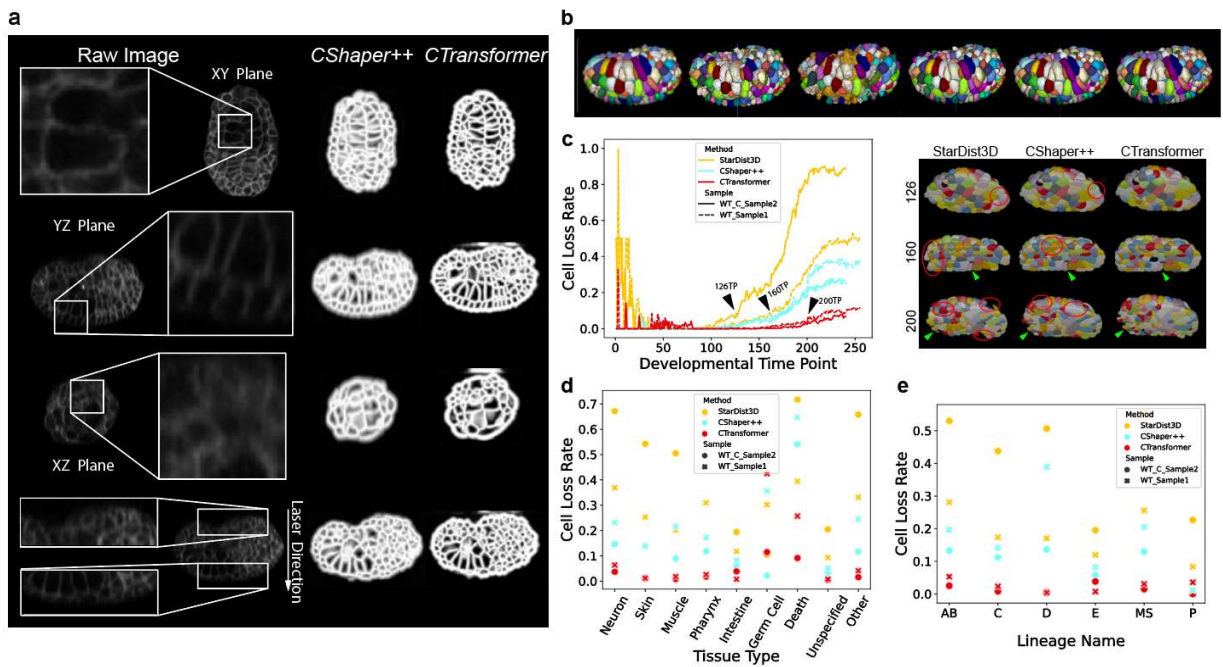

**Fig 4 | Overall 3D cell membrane tracking performance in terms of image enhancement, cell morphology reconstruction, and cell lineage tracing.** (**a**) Image enhancement at ~550-cell stage. While cell blurring occurs in different regions due to different reasons (e.g., at the axial imaging plane, the first and fourth rows, ambiguous areas are prevalent because of isotropic resolution, light scattering, and low resolution), CTransformer enhances the image quality substantially and better than CShaper++. (**b**) Cell morphology reconstruction at different developmental stages. (**c**) Cell loss rate in compressed and uncompressed embryo samples, regarding developmental time. (**d**) Cell loss rate in compressed and uncompressed embryo samples, regarding cell fates. (**e**) Cell loss rate in compressed and uncompressed embryo samples, regarding cell lineages.

**With the *m2nGAN* module, *CTransformer* traces cell lineages accurately by generating pseudo cell nuclei fluorescence images near the real ones.**

The first module of *CTransformer* is *m2nGAN*: a membrane-to-nucleus generative adversarial network that synthesizes high-fidelity pseudo-nucleus images from membrane-only 4D images, thereby enabling nucleus-free lineage tracing (Fig. 1c). The architecture comprises dual generators ($G_A$, $G_B$) and dual discriminators ($D_A$, $D_B$) that are jointly optimized to reconstruct nucleus morphologies with high contrast and biological plausibility. Generator $G_A$ is conditioned on the precise cell boundaries provided by **4dTUNETr** and outputs synthetic nuclei that surpass conventional nucleus labels in signal clarity, reducing background noise and improving the visibility of subnucleus structures (Figure. 1b and S5). This approach not only bypasses the need for nucleus labeling in live imaging but also enhances lineage reconstruction accuracy in densely packed cellular environments.

To quantitatively evaluate the pseudo generative (results of *m2nGAN*) and real raw nuclei images (second row in Figure 1c), 2 image-wise metrics, peak signal to noise ratio (PSNR) based on the mean square error (MSE) and signal-to-noise ratio (SNR) were used, as well as the standard criteria, precision, recall, and accuracy, on the "**Lineage Tracing Evaluation**" dataset. This dataset includes 3 time series, 1 compressed and 2 uncompressed of embryos with 625 embryonic volumes, as well as all their edited nuclei (Figure S6); all of their maximum imaging time points are up to the 550-cell stage. The effectiveness of *m2nGAN* at constructing generative pseudo nuclei images to establish cell lineages is demonstrated (Figure 3).

PSNR is defined as

$$PSNR = 20 \times log_{10}\left(\frac{MAX_f}{\sqrt{MSE}}\right), \quad (4)$$

where $MAX_f$ is 255 and the MSE is formulated as $MSE = \frac{1}{mn}\sum_{i=1}^{m}\sum_{j=1}^{n}\left(I_{gt}(i,j), I(i,j)\right)^2$.

SNR is calculated as $SNR = \frac{P_{signal}}{P_{noise}}$, defined for images of nuclei measured in this study. $P_{signal}$ and $P_{noise}$ represent the total power of signals and noise, respectively, as measured by the accumulated intensity of the pixels inside and outside the nuclei. Signal refers to the voxels located within the cell nuclei regions identified in GT images and noise consists of voxels outside those regions.

Precision, recall and accuracy are formulated as:

$$Precision = \frac{True\ Positives}{True\ Positives + False\ Positives}, \quad (5)$$

$$Recall = \frac{True\ Positives}{True\ Positives + False\ Negatives}, \quad (6)$$

$$Accuracy = \frac{True\ Positives}{True\ Positives + False\ Positives + False\ Negatives}. \quad (7)$$

At every time point, they are computed according to the cell identities within the time series (lineage tree), as has been done widely previously[35-37]. Error numbers are compared and categorized into 5 types (Figure 3b), i.e., division error and tracing error. Qualitative assessments of the whole lineages and single imaging slides (intensity heterogeneity) and a comprehensive lineage-wise evaluation of the generative images, from the perspective of tracing of the cell lineage by biologists are also provided.

Additionally, qualitative traced lineage tree comparisons between raw (real fluorescence) and generative images also demonstrated the effectiveness of *m2nGAN* (Figures S9 to S11), highlighting the following 3 advantages of *m2nGAN*,

(1) In top, middle, and bottom focal planes, pseudo-nucleus exhibited stronger and more uniform signals, with reduced background artifacts.

(2) In early to late embryonic stages, pseudo-nucleus images maintained consistent contrast and structural clarity (Figure 3e), supporting stable lineage inference over time.

(3) Lineage trees constructed from generative images were completer and more consistent with biological expectations.

For a more comprehensive evaluation, we conducted the quantitative **image-wise** and standard measurements to show that *m2nGAN* generated images of nuclei are better than real (raw) images they have higher accuracy and SNR, lower MSE and error numbers than real nuclei images (Figures 3a, S6, and S7) for cell lineaging. Construction and manual curation of cell lineages became easier and more efficient with fewer tracking and tracing errors in generative images (Figure 3b, Figure S8).

With above evaluated results, generated images (the results of *m2nGAN*), give equally good or even better signals of nuclei and image quality to allow lineage tracing to build cell morphological maps.

As The *m2nGAN* Module allows cell nucleus position remodeling and lineage based on only the cell membranes, dynamics of other molecules of interest, not only for the previously nuclei-relied ones (e.g., promotor and chromatin activities) but also for the ones displaying spatial dynamics beyond nucleus (e.g., HMR-1/Ecadherin located on cell membrane surface/interface), can be monitored with additional fluorescence channel on top on the cell morphology and cell lineage

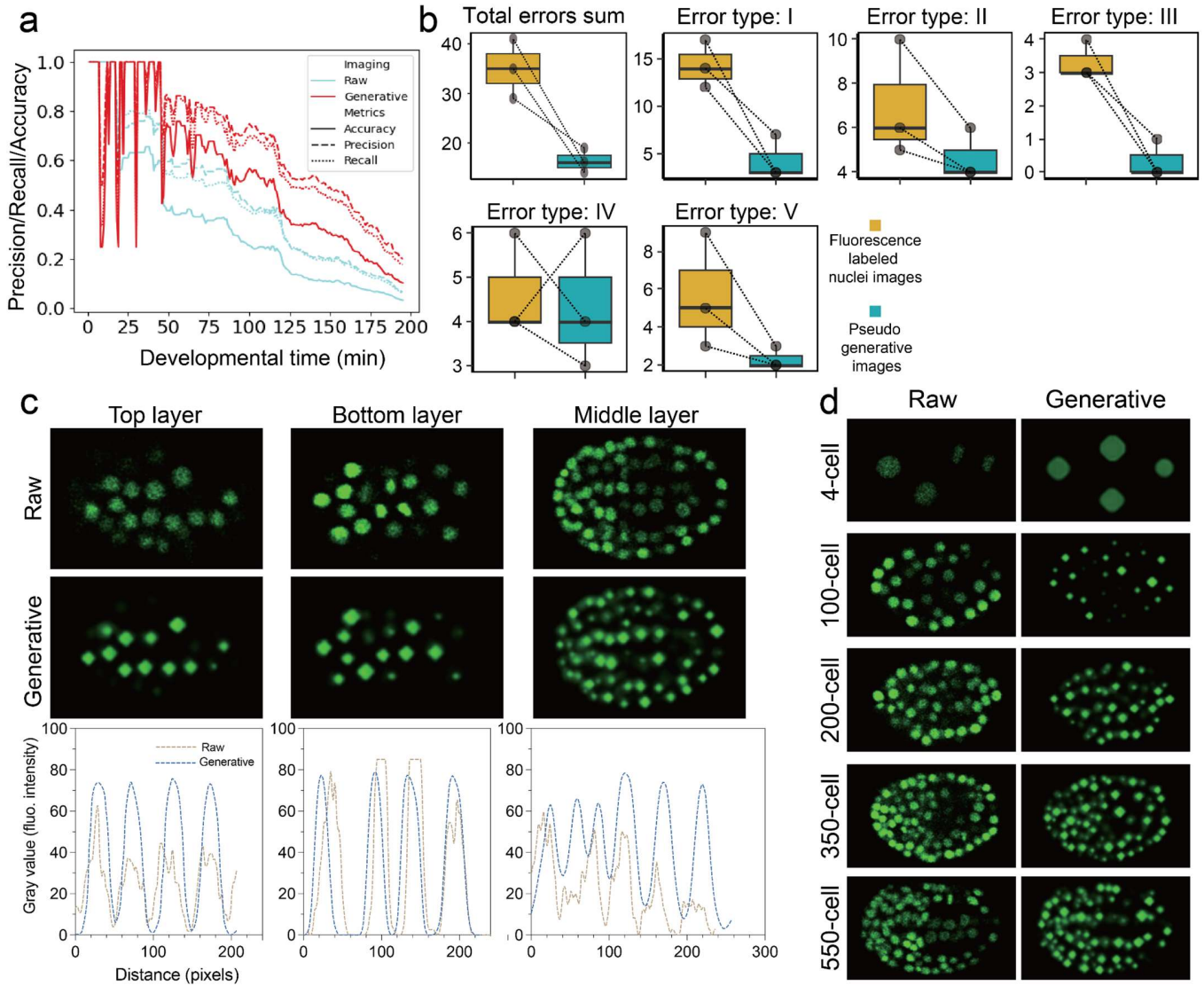

**Fig 3 | Comparable real (based on cell nuclei fluorescence) and generative (based on cell membrane fluorescence and pre-trained *m2nGAN* module) images for cell lineage tracing.** (**a**) The precision (dashed line), recall (solid line), and accuracy (dotted line) curves of cell lineage tracing with real (blue) and generative (red) images. (**b**) The number of errors produced by real (yellow) and generative (cyan) image, measured at the ~160-cell stage from three individual embryos samples. A dashed line connects the error counts for the statistical results in each embryo of raw (real) and generative (pseudo) images. (**c**) Smoother, more even and more distinct image signal distribution in generative images than real images, displayed with different focal planes The intensities of fluorescence of raw (blue) and generative (red) images, measured in the whole image. (**d**) Smoother, more even and more distinct image signal distribution in generative images than real images, displayed with different developmental stages.

**With the *m2nGAN* Module, *CTransformer* automatically quantifies spatiotemporal dynamics of a labeled molecule across multicellular to subcellular scales with the spare channel.**

The third component of the *CTransformer* pipeline consists of a set of specialized algorithms designed for **shape-based representation analysis**, which play a critical role in interpreting complex 4D cellular morphology. These algorithms extract high-dimensional geometric features from segmented cell membranes, including shape descriptors, spatial orientation, and relative positioning within the embryo context. By analyzing how these shape features evolve over time and across lineages, the pipeline enables the identification of developmental patterns such as coordinated cell elongation, directional growth, and spatial organization of cell assemblies. Furthermore, by integrating adhesion information from the second imaging channel, the system can correlate morphological changes with cell–cell contact dynamics, offering insights into how physical interactions drive morphogenetic processes.

*CTransformer*'s remarkable capability to track 3D cell membranes independently of cell nuclei offers a chance to quantify spatiotemporal dynamics of any molecule of interest through fluorescent labeling in a separate channel (Figures 2, 3, 4). This capability is technically important because the number of distinct fluorescent markers and non-overlapping imaging channels in microscopy is often limited. Since cell morphologies and cell lineages across the entire organism can be reconstructed solely from cell membrane fluorescence, such quantification can reach subcellular resolution. In other words, this methodology allows monitoring how molecular distributions vary within and across 3D cell bodies over time accurately. To demonstrate this idea, we utilized the protein HMR-1/E-cadherin, which is well-known to accumulates at cell interfaces and mediates adhesion between neighboring cells; As an evolutionarily conserved protein from worms to humans, it has been implicated in multiple developmental processes, including body axis establishment, germ layer segregation, and intestinal formation, and other key morphogenetic processes. With fluorescent intensity values are used as a proxy for the physical surface adhesion in the following quantitative analyses. In metazoan development, induction of cellular fates via Wnt and Notch signaling[39-41], often relies on continuous and direct contact and adhesion between specific cells to ensure proper functional interactions[12, 42, 43], for instance, interactions between ligands and their receptors to initiate signaling pathways[10, 11, 44]. By using digitized and detailed morphological data provided in this study, researchers can infer these cellular interactions by examining factors like the area and duration of cell contact. These cell-to-cell interactions and their adhesion level play a pivotal role in triggering cellular changes. The reproducibility of adhesion in cell-cell contacts was examined with 3 time-lapse samples involving 3613 contact pairs over 40008 time points with quantitative measurements of hmm-1 fluorescence. This showed that a moderate level of variability was tolerated in cell-cell contacts with 64% of data points having variation ratio lower 0.3 (Figure 5e). We demonstrated that cell-cell interaction adhesions controlled by remarkable invariances and reproducibilities for precise cellular development, like cell volume, surface area, contact area in previous studies[6, 45]. The 5 contact pairs with highest hmm-1 fluorescence over at least 3 consistent time points (Figure 5f), except for ABprpaaa, are skin and intestine cells, as, with muscle cells, are over 80% of cells with high inferred adhesion levels. This might reflect the need to maintain a tighter, fixed shape.

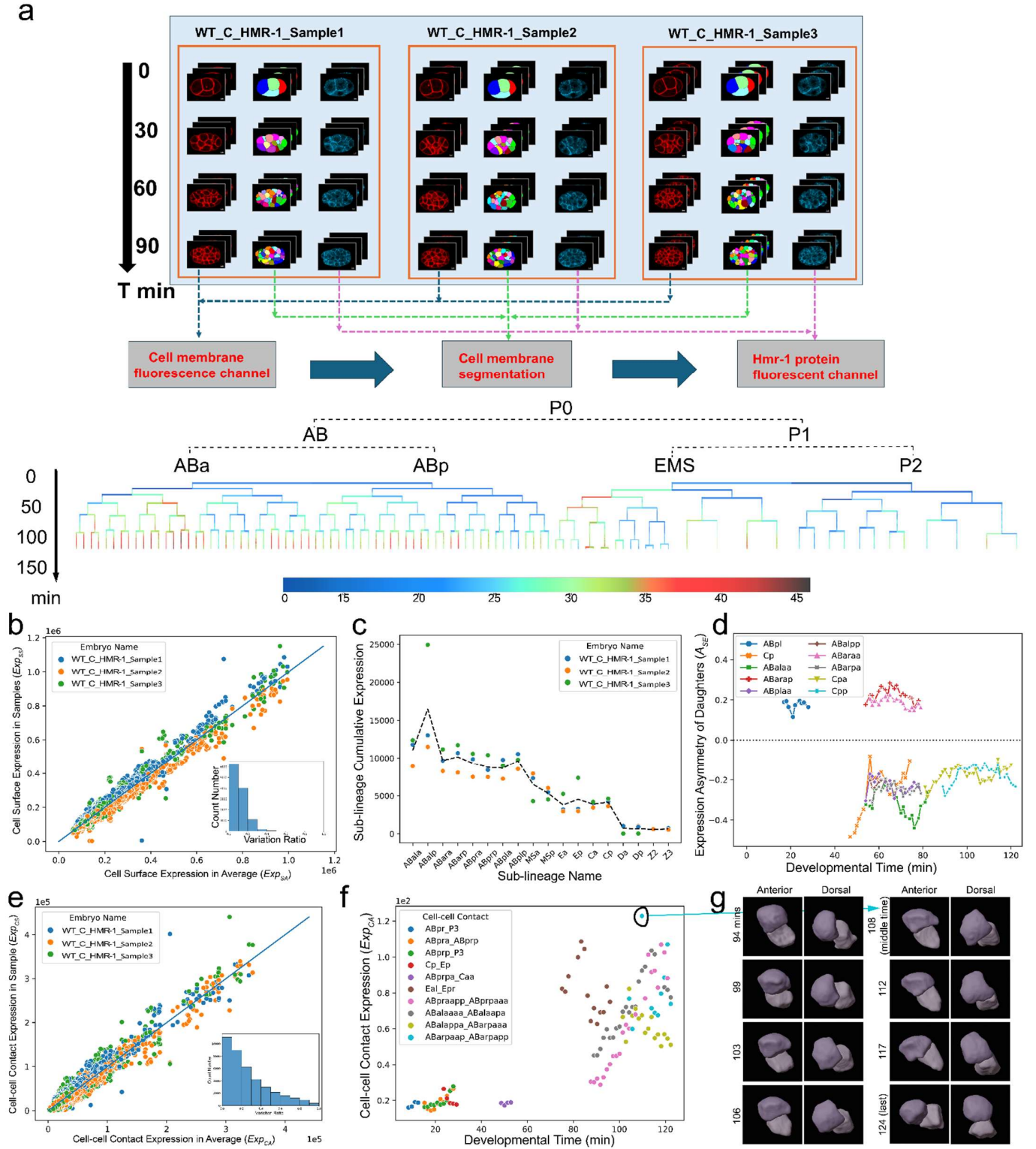

**Fig 5. Whole-organism and subcellular-resolved quantification of HMR-1/E-cadherin dynamics governing adhesion.** (**a**) Cell adhesion, expressed in units (0.0324 $\mu m^2$) of surface area, over a complete lineage tree. (**b**) Precision of cellular adhesion across their complete existing durations among three embryo samples. Inset: histogram of variation coefficients. (**c**) The cumulative expression of surface adhesion of the 18 spatial cell sub-lineages. (**d**) Cellular adhesion asymmetry between daughters of the ten cells with highest

adhesion (presented in time-lapse curves), revealing persistent adhesion asymmetry with alternating directional biases. (**e**) Precision of subcellular adhesion across their complete existing durations among three embryo samples. Inset: histogram of variation coefficients. (**f**) Subcellular adhesion of the five contact pairs with strongest adhesion and five contact pairs with weakest (normalized absolute value) (presented in time-lapse curves), revealing persistent, fluctuating, or trending adhesion over time. (**g**) Interacting cells with the largest expression contact area over their lifecycle: ABarpaap_ABarpapp.

Cell adhesion, represented by HMR-1/E-Cadherin in C. elegans, is known to play a conserved role in different biological conditions spanning from embryonic development and immunal response, and spanning from low-complexity organism like worm to high-complexity like human. Although observation is usually with subcellular resolution to highlight relative distribution within the cell body, most experimental reports show a collective dynamics such as a global change or an axis dependent change, ignoring how cell adhesion may vary upon specific cells and cell-cell contacts. Nevertheless, previous C. elegans experiment shows heterogeneous cell adhesion at its 4-cell stage, which controls the diamond patterns critical for Notch and Wnt signaling and implies a mechanism of conversation pool of such cell adhesion protein. To the best of our knowledge, whether a cell-membrane-bound molecule can be regulated down to subcellular regulation over dozens to hundreds of cells is unclear, which represents a biophysical limit of developmental control.

Precise adhesion control from the cellular to subcellular scales exhibits specificity across lineage, spatial, and temporal, and dimensions. In the lineage scale, the AB lineage exhibits stronger adhesion than P1, which may account for previous observations that AB cells are more regular than P1 cells. While the AB lineage occupies the anterior and the P1 lineage the posterior, the particularly stronger adhesion (compared to ABp sublineage) observed in the more anterior ABa sublineage within AB lineage further; This anterior preference, observed across all 18 sublineages (Figure 5b), further suggests that the spatial heterogeneity in adhesion is biased along the anterior-posterior axis as a global manner. Whether this results from internal lineage differentiation or external diffusive signaling (*e.g.*, Wnt signaling) warrants further experimental investigation. In consideration of the stereotypical bias of cell fate in AB and P1 lineages, whether these higher surface adhesion areas contribute to the further differentiation of pharyngeal and neuronal tissues concentrated in AB lineage, or in contrary, low adhesion contributes to differentiation in P1 lineage, is an interesting developmental biology and biophysical question worth future investigation (Figure 5a).

Down to the cellular scale, the ten cells with highest adhesion exhibit adhesion asymmetry between their daughter cells; however, whether the anterior or posterior daughter inherits the higher adhesion appears to be cell-dependent in a local manner (Figure 5a). For 11583 cell measurements in 3 embryos, the cell adhesion surfaces were generally invariable with over 90% of cells having a variation level lower than 0.2 ratio (Figure 5b), indicating a high level of reproducibility of adhesion surfaces in developing embryos, which might drive anterior-posterior, left-right, and dorsal-ventral development (Figure 5c). Directional asymmetries (see Supplementary Methods) of daughter cell adhesion patterns may relate to differentiation (Figure 5d). ABpl, ABaraa, and ABarap showed positive asymmetries (the anterior daughter cell showed higher adhesion related fluorescence than the posterior cell). Cp, Cpa and Cpp differentiate as skin and muscle (see Cell Fate xslx). These 2 daughter cells and their antecedents have high negative asymmetries and accurately move to form skin and muscle tissues, covering and supporting the entire embryo. The high asymmetric of fluorescent signals of ABalaa and ABalpp (which differentiate to neuronal and apoptotic tissues, respectively), and ABarpa and ABplaa (which differentiate to neuronal, skin, and death tissues), (Figure 5d).

## Discussion

Since about two decades ago, fluorescence labeling on different biological parts have been a widely used way to extract information, especially in the 3D + t scenario where numerous data is generated with standard format. Typically, cell nucleus labeling allows cell lineage tracing considering its consensed signal and high contrast with respect to a black background, whereas cell membrane labeling enables complete cell morphology reconstruction. While cell nuclei cannot provide the complete information of cell membrane, cell membrane as the complete cell body in space may cover the the needed information for cell lineage tracing that usually depend on cell nucleus, which is a point inside the cell body. Inspired by this fact, this study devise an experimental-computational framework, CTransformer, to automatically reconstruct cell morphology and trace cell lineage with only one fluorescence channel for cell membrane labeling, bypassing laborous human intervention required before (Figure 1). Demonstrated with the worm C. elegans, CTransformer's *TUNETr Module* achieve significantly outperform other state-of-the-art algorithms throughout embryogenesis up to a crowded stage with over 550 cells within its eggshell, in terms of both high segmentation accuracy (Figure 2) and low cell loss (Figure 3). *Since* the cell morphology reconstruction overcomes previous failure in cells with extremely deformed shapes (e.g. skin cells), and extremely small sizes (e.g. apoptotic cells), (whose membrane signals were too weak and blurry), CTransformer's *m2nGAN Module* is able to make use of them for generating pseudo-nuclei images and cell lineage tracing (Figure 4). Such accurate cell morphology reconstruction reduces the potential for an expanding chain errors in cell lineage tracing, jointly spare the original cell nucleus channel so that CTransformer's *MolQuantifier Module* can use it to label another molecule of interest, exemplified by the well-known and critical adhesive protein HMR-1/E-Cadherin (Figure 5). Quantitative labeling demonstrate previously unknown features of precisely regulated adhesion of specific cells and cell-cell contacts (Figure 5). By full utilization of image information and enabling subcellual-resolution molecular quantification, CTransformer is reasonably applied for a broad range of scenarios.

1. The spared channel(s) can accommodate labeling not only on the HMR-1/E-Cadherin, but also any other labelable molecule. When modulating fluorescence colors and applying three and more colors markers, dynamics of two or more molecules can be jointly quantified. This permit the landscape scanning beyond the previous dimensions including but not limited to gene expression, chromatin accessibility, and data with different dimensions can be studied as a whole.
2. Cell properties such as adhesion (labeled by HMR-1/E-Cadherin), tension (labeled by F-actin), and others can be estimated without harming the sample. Compared with other unharmful indirect estimation methods represented by image-based mechanical modeling/calculation, CTransformer set up the focus on molecular labeling, which is closer to the reality since the force interpreted is directly attributed to a certain molecule without unambiguity. Conversely, indirect estimation methods, even the inferred force is correct, it represents a resultant effect and misses the detailed information such as the origins this force comes from.
3. The general design of CTransformer unleash its applicability not only on the exemplary C. elegans embryo, but also other system with comparable level of cell crowdness. For example, ascidian system is traced with its first half of embryogeness, aided with a large amount of manual tracing. CTransformer could be applied in such scenario and also the systems with abnormal condition (RNAi, mutation, etc), as long as there is a cell membrane fluorescence channel. Importantly, the reliance of only cell membrane fluorescence but not nucleus fluorescence mean that the 3D +t imaging can have roughly half of photobleaching and phototoxicity, and equavality, allowing double duration for the sample to image.

The whole CTranformer framework is applied to both previously published and newly added C. elegans embryos, providing an informative resource including XXX embryos for both developmental biology and computer science research. Dated back to the developmental path of cell nucleus tracing of StarryNite/AceTree to gene expression profiling of AceBatch, followed by the development path of cell morphology reconstruction algorithms from 3DMMS, CShaper, CMap, and CTransformer, we anticipate the new era of measuring molecular distribution on top the previous both.

# Methods

### *C. elegans* strains

All animals were maintained on nematode growth media (NGM) plates seeded with *Escherichia coli* (OP50) at room temperature. First, for embryos subjected to nucleus-based cell lineage tracing and membrane-based cell morphology segmentation, strains ZZY0861 and ZZY0655 were used. The transgenic strain RB669 (wee-1.1(ok418)) was crossed with the strain ZZY0861, resulting in a new strain ZZY1142 with both *wee-1.1* deficiency and fluorescence labeling on cell nucleus and membrane. Second, a cell-membrane-labeled strain ZZY0637 (carrying a single copy of PH2::mCherry) was crossed with the HMR-1-labeled strain LP172 (carrying a GFP-tagged endogenous *hmr-1* gene). Both markers were rendered homozygous in the resulting strain ZZY0868 (hmr-1(cp21[hmr-1::GFP + LoxP]) I; unc-119(tm4063) III; zzySi139 [Phis-72::PH(PLC1delta1)::mCherry::pie-1 3' UTR + unc-119(+)] II).

## 3D+$t$ fluorescence imaging

For *C. elegans* embryos with both fluorescence labeling on cell nucleus (GFP) and cell membrane (mCherry), the imaging protocol followed our previous studies. Fluorescence images start to be captured in embryos with 4 cells at most using either the SP5 II or Stellaris confocal microscopy system from Leica. Then, they are processed and analyzed using the LAS X software from Leica and ImageJ. To keep the embryos alive, the intensity of light was set to be low (a scanning speed of 8,000 Hz with a water immersion objective). The imaging process was separated into successive time blocks of 60 time points. The *z*-axis compensation ranged from 0.5-3% for the 488-nm laser and 20-95% for the 594-nm laser. Subsequently, *StarryNite* and *AceTree*[33, 35, 49] were used to recognize every cell nucleus (cell index: $i$) at consecutive time points (time index: $t$) in each individual embryo (embryo index: $j$), systematically recording their 3D positions $(x_{i,j,t}, y_{i,j,t}, z_{i,j,t})$.

For *C. elegans* embryos with both fluorescence labeling on HMR-1 (GFP) and cell membrane (mCherry), the imaging protocol followed a combination of methods described in our previous studies, as detailed here. One to four-celled embryos were mounted for imaging using 1% methylcellulose in Boyd's buffer with 20 μm Polybead® microspheres (Polysciences, Inc.); fluorescence imaging was performed with an inverted Leica SP5 confocal microscope equipped with two hybrid detectors at a constant ambient temperature of 21 °C. Images were simultaneously collected for both GFP and mCherry channels using a water immersion objective; both channels were imaged with a frame size of 712 × 512 pixels (0.09 μm/pixel) combined with a scanning speed of 8,000 Hz using a resonance scanner. The excitation laser beams used for GFP and mCherry were 488 nm and 594 nm respectively; images containing 68 *z*-steps were collected for three individual embryos with a *z*-resolution of 0.42 μm from top to bottom for every time point, which was at 1.41-minute intervals. Images were continuously collected for 100 time points with the embryos developing into "E4" ("E", the intestine progenitor, dividing into four granddaughters) or "E8" stages ("E", the intestine progenitor, dividing into four granddaughters). The *z*-axis compensation was 0.5-8% for the 488-nm laser and 19-95% for the 594-nm laser with the pinhole size of 2.5 AU (airy unit). Later, a deconvolution process was used for fluorescence images before the application of *CTransformer* framework.

## *4dTUNETr*

We improved the input and output of the DNN module of *TUNETr* (Figures S1 and S2, see Supplementary Information), as well as the cell instance segmentation for a better and more robust single-cell segmentation for pseudo nuclei images generations. We give clear nuclei sizes and positional hints (both from segmented single-cell morphology) for GAN to generate the nuclei with relatively accurate nuclei fluorescent signals (Figure S16). Thus, *TUNETr* was trained with only the membrane channel (*iTUNETr* and *sTUNETr* used the nucleus channel for training and running). To generate stable recognition of membranes, the 4D images from the synthetic dataset were used for training in semi-supervised learning *sTUNETr* (see steps 2 and 3 of Section Detailed Training and Running Segmentation Workflow in Supplemetary Materials). For the 3 input sequential volumes, *T-1*, *T*, *T+1*, the output would be predictions of these 3 time points. *T-1* and *T+1* would also generate *T-2*, *T-1*, *T*, and, *T*, *T+1*, *T+2*, respectively. As for a shifted window, at time point *T*, *4dTUNETr* produced 3 membrane predictions *T'*, *T''*, and *T'''*. The binarization process takes advantage of these predictions to integrate a robust membrane segmentation. In one of the evaluated embryos, *4dTUNETr* produces 524 while *sTUNETr* produces 506 cells for 550-cell embryo.

## *m2nGAN*: The GAN and Generative Network for Fluorescent Pseudo Nuclei Images

Segmented single-cell results from *4dTUNETr* (improved *sTUNETr*) were fed to the 3D Cycle Generative adversarial network (3D−cycle−GAN) to generate time−lapse pseudo−nuclei fluorescence images[50, 51]. Only membrane−labelled 4D images (*4dTUNETr*) are needed to build the cell−resolved morphology map for living organisms.

With single−cell segmentations from *4dTUNETr*, an octahedron is artificially generated for every cell as hints for *m2nGAN*. These octahedra are produced by dilation operations with N iterations where N is derived from the R of one cell, formulated as

$$N = min[n \in Z \mid x \leq R].$$

And the R is calculated as

$$R = \sqrt[3]{\frac{3 \times V_{voxel}}{4 \times \pi}}.$$

The voxels of localized and size-fixed octahedra were set as 255 in the 8-bit volumetric images. With multiple octahedra in a 3D volume, *m2nGAN* can transform the distributions of images to the real nuclei fluorescence images.

## *MolQuantifier*

We develop *MolQuantifier* simultaneously to quantify the modular distributions on every single-cell. MolQuantifier could measure the RNA expression imaged in the fluorescence images. On every segmented cell surface (it consists of three layers of voxels, simulating the real structure of cell membranes), MolQuantifier calculate all the expression values on the membrane. This quantification method can accurately obtain expression on the 3D selected area, like membrane, nucleus, cytoplasm, contacting surface, rather than on the only estimated whole cell area.

## Data and code availability

All data and code are accessible via https://doi.org/10.6084/m9.figshare.27085657 or will be provided upon reasonable request.

# Acknowledgements

We thank Prof. Xiaojing Yang (Peking University) and Fengxi Yu (Tsinghua University) for their invaluable assistance in initiating this project. This work is supported by Hong Kong Innovation and Technology Commission (InnoHK Project CIMDA) and Hong Kong Research Grants Council (Project 11204821) to H.Y., and by Hong Kong Innovation and Technology Commission (GHP/176/21SZ) and Hong Kong Research Grants Council (Projects HKBU12101520, HKBU12101522, HKBU12101323) to Z.Z.

# References

1. Fujii, Y. et al. Spatiotemporal dynamics of single cell stiffness in the early developing ascidian chordate embryo. *Communications Biology* **4** (2021).
2. Brangwynne, C.P. et al. Germline P Granules Are Liquid Droplets That Localize by Controlled Dissolution/Condensation. *Science* **324** (2009).
3. Otsuji, M. et al. A mass conserved reaction-diffusion system captures properties of cell polarity - PubMed. *PLoS computational biology* **3** (2007).
4. Stegmaier, J. et al. Real-Time Three-Dimensional Cell Segmentation in Large-Scale Microscopy Data of Developing Embryos. *Developmental Cell* **36**, 225-240 (2016).
5. Azuma, Y. & Onami, S. Biologically constrained optimization based cell membrane segmentation in C. elegans embryos. *BMC Bioinformatics* **18** (2017).
6. Cao, J. et al. Establishment of a morphological atlas of the *Caenorhabditis elegans* embryo using deep-learning-based 4D segmentation. *Nature Communications* **11** (2020).
7. Wang, F. et al. Far-field super-resolution ghost imaging with a deep neural network constraint. *Light: Science & Applications* **11** (2022).
8. Azuma, Y. & Onami, S. Biologically constrained optimization based cell membrane segmentation in *C. elegans* embryos. *BMC Bioinformatics* **18**, 307 (2017).
9. Waliman, M. et al.  (Cold Spring Harbor Laboratory, 2024).
10. Chen, L. et al. Establishment of Signaling Interactions with Cellular Resolution for Every Cell Cycle of Embryogenesis. *Genetics* **209**, 37-49 (2018).
11. Ma, X. et al. A 4D single-cell protein atlas of transcription factors delineates spatiotemporal patterning during embryogenesis. *Nature Methods 2021 18:8* **18** (2021).
12. Guan, G. et al.  (Cold Spring Harbor Laboratory, 2023).
13. Ho, V.W.S. et al. Systems-level quantification of division timing reveals a common genetic architecture controlling asynchrony and fate asymmetry. *Molecular Systems Biology* **11**, 814 (2015).

14. Maître, J.-L. & Heisenberg, C.-P. Three Functions of Cadherins in Cell Adhesion. *Current Biology* **23**, R626-R633 (2013).
15. Calcutt, R., Vincent, R., Dean, D., Arinzeh, T.L. & Dixit, R. Plant cell adhesion and growth on artificial fibrous scaffolds as an in vitro model for plant development. *Science Advances* **7** (2021).
16. Guignard, L. et al. Contact area–dependent cell communication and the morphological invariance of ascidian embryogenesis. *Science* **369**, eaar5663 (2020).
17. Cao, J., Wong, M.-K., Zhao, Z. & Yan, H. 3DMMS: robust 3D Membrane Morphological Segmentation of C. elegans embryo. *BMC Bioinformatics* **20** (2019).
18. Cao, J. et al. CShaperApp: Segmenting and analyzing cellular morphologies of the developing Caenorhabditis elegans embryo. *Quantitative Biology* (2024).
19. Qiu, C. et al. A single-cell time-lapse of mouse prenatal development from gastrula to birth. *Nature 2024 626:8001* **626**, 1084-1093 (2024).
20. Heymann, J.A.W. et al. Site-specific 3D imaging of cells and tissues with a dual beam microscope. *Journal of Structural Biology* **155**, 63-73 (2006).
21. Abouakil, F. et al. An adaptive microscope for the imaging of biological surfaces. *Light: Science & Applications* **10** (2021).
22. Park, H. et al. Deep learning enables reference-free isotropic super-resolution for volumetric fluorescence microscopy. *Nature Communications* **13** (2022).
23. Tahir, W., Wang, H. & Tian, L. Adaptive 3D descattering with a dynamic synthesis network. *Light: Science & Applications* **11** (2022).
24. Ning, K. et al. Deep self-learning enables fast, high-fidelity isotropic resolution restoration for volumetric fluorescence microscopy. *Light: Science & Applications* **12** (2023).
25. Li, X. et al. Spatial redundancy transformer for self-supervised fluorescence image denoising. *Nature Computational Science* **3**, 1067-1080 (2023).
26. Schmidt, U., Weigert, M., Broaddus, C. & Myers, G. in Medical Image Computing and Computer Assisted Intervention – MICCAI 2018 265-273 (Springer International Publishing, 2018).
27. Stringer, C., Wang, T., Michaelos, M. & Pachitariu, M. Cellpose: a generalist algorithm for cellular segmentation. *Nature Methods* **18**, 100-106 (2021).
28. Wang, A. et al. A novel deep learning-based 3D cell segmentation framework for future image-based disease detection. *Scientific Reports* **12** (2022).
29. Vaswani, A. et al. Attention is all you need. *Advances in neural information processing systems* **30** (2017).
30. Panaretos, V.M. & Zemel, Y. Statistical Aspects of Wasserstein Distances. *Annual Review of Statistics and Its Application* **6**, 405-431 (2019).

31. Clough, J.R. et al. A Topological Loss Function for Deep-Learning Based Image Segmentation Using Persistent Homology. *IEEE Transactions on Pattern Analysis and Machine Intelligence* **44**, 8766-8778 (2022).
32. Rasmussen, J.P., Feldman, J.L., Reddy, S.S. & Priess, J.R. Cell Interactions and Patterned Intercalations Shape and Link Epithelial Tubes in C. elegans. *PLoS Genetics* **9**, e1003772 (2013).
33. Murray, J.I. et al. Automated analysis of embryonic gene expression with cellular resolution in C. elegans. *Nature Methods* **5**, 703-709 (2008).
34. Malin-Mayor, C. et al. Automated reconstruction of whole-embryo cell lineages by learning from sparse annotations. *Nature Biotechnology 2022 41:1* **41** (2022).
35. Bao, Z. et al. Automated cell lineage tracing in Caenorhabditis elegans. *Proceedings of the National Academy of Sciences* **103**, 2707-2712 (2006).
36. Santella, A., Du, Z., Nowotschin, S., Hadjantonakis, A.-K. & Bao, Z. A hybrid blob-slice model for accurate and efficient detection of fluorescence labeled nuclei in 3D. *BMC Bioinformatics* **11**, 580 (2010).
37. Santella, A., Du, Z. & Bao, Z. A semi-local neighborhood-based framework for probabilistic cell lineage tracing. *BMC Bioinformatics* **15**, 217 (2014).
38. Andrew D., C. & Jeff, H. Epidermal morphogenesis. (*WormBook*, 2005).
39. Walston, T. et al. Multiple Wnt Signaling Pathways Converge to Orient the Mitotic Spindle in Early C. elegans Embryos. *Developmental Cell* **7**, 831-841 (2004).
40. Thorpe, C.J., Schlesinger, A., Carter, J.C. & Bowerman, B. Wnt Signaling Polarizes an Early C. elegans Blastomere to Distinguish Endoderm from Mesoderm. *Cell* **90** (1997).
41. Neves, A. & Priess, J.R. The REF-1 Family of bHLH Transcription Factors Pattern C. elegans Embryos through Notch-Dependent and Notch-Independent Pathways. *Developmental Cell* **8**, 867-879 (2005).
42. Pocock, R., Bénard, C.Y., Shapiro, L. & Hobert, O. Functional dissection of the C. elegans cell adhesion molecule SAX-7, a homologue of human L1. *Molecular and Cellular Neuroscience* **37**, 56-68 (2008).
43. Asan, A., Raiders, S.A. & Priess, J.R. Morphogenesis of the C. elegans Intestine Involves Axon Guidance Genes. *PLOS Genetics* **12**, e1006077 (2016).
44. Du, Z. et al. The Regulatory Landscape of Lineage Differentiation in a Metazoan Embryo. *Developmental Cell* **34**, 592-607 (2015).
45. Li, Z., Cao, J., Zhao, Z. & Yan, H. (Research Square Platform LLC, 2021).
46. *Jayesh, V.*, *Arulkumar, S.* & A*nurag,* M. in Proceedings of the IEEE/CVF winter conference on applications of computer vision (CVPR) 2774-2784 (2022).
47. Guanxiong, S., Yang, H., Guosheng, H. & Neil, R. in Proceedings of the AAAI Conference on Artificial Intelligence, Vol. 35 (2021).

48. Albert, G. & Tri, D. Mamba: Linear-Time Sequence Modeling with Selective State Spaces. *arXiv preprint* (2023).
49. Boyle, T.J. et al. AceTree: a tool for visual analysis of Caenorhabditis elegans embryogenesis. *BMC Bioinformatics 2006 7:1* **7** (2006).
50. Ge, Y. et al. in 2019 IEEE 16th International Symposium on Biomedical Imaging (ISBI 2019) (IEEE, 2019).
51. Isola, P., Zhu, J.-Y., Zhou, T. & Efros, A.A. in IEEE Conference on Computer Vision and Pattern Recognition (CVPR) 1125-1134 (2017).

# Supplementary Information of *CTransformer*

## Materials and Methods

**The Animal *C. elegans*, Cell Morphology Map and the Embryogenesis**

The model organism *Caenorhabditis elegans*, *C. elegans*, is widely used to study cellular developmental biology during embryogenesis due to its highly transparent body and regular cellular development, including consistent patterns in cell lineages (Sulston et al. 1983), timings of divisions (Fickentscher, Struntz and Weiss 2016), orientations of axes (Sugioka and Bowerman 2018), and gene expression (Ma et al. 2021, Boyle et al. 2006, Murray et al. 2008). To track these cellular processes and patterns, previous studies used transgenic markers, like GFP or *mCherry*, to trace lineages and images of fluorescent nuclei to show cellular differentiation, migration and interactions (Bao et al. 2006, Kretzschmar and Watt 2012, Mcdole et al. 2018, Malin-Mayor et al. 2022, Fire 1994, Sugawara, Çevrim and Averof 2022). The fluorescent labels allow descendants of individual cells to be tracked, with the lineages showing the cell cycles of developmental patterns (Fickentscher et al. 2016, Guan et al. 2021, Cao et al. 2020, Guan et al. 2023, Ma et al. 2021).

Linking cell morphology to cell identity during *in vivo* embryogenesis allows investigation of the regulation of cell-cell contacts, gene expression and determination of cell fates during development (Rasmussen et al. 2013, Cao et al. 2020, Goldstein and Nance 2020, Pentinmikko et al. 2022). Four-dimensional (4D), 3D + time (t), imaging is the basic technique that could achieve this by construction of a series of time-lapse images of embryonic development showing cellular morphology and cell lineages (Domcke and Shendure 2023, Qiu et al. 2024, Packer et al. 2019).

Multicellular animals, metazoans, show a range of complex deformations of cellular morphology (Guan et al. 2023, Chen et al. 2018, Mittenzweig et al. 2021, Junyent et al. 2024) and the durations of cell cycles are highly controlled by the asymmetry in sizes of single-cells (Guan et al. 2021, Fickentscher, Struntz and Weiss 2016, Li, Zhao and Yan 2023b, Jankele et al. 2021), with cell junctions and adhesions playing important developmental roles(Maître and Heisenberg 2013, Lee et al. 2014, Asan, Raiders and Priess 2016). Specification of cell fate(Mello, Draper and Priess 1994), programmed migration(Li et al. 2019), genetic regulation of cell shape(Guan et al. 2023, Maître and Heisenberg 2013), and cellular contacts(Ho et al. 2015, Cao et al. 2020, Chen et al. 2018) can all benefit from a morphological map. Tracing complete cell lineages needs images of fluorescently stained nuclei (Sugawara, Çevrim and Averof 2022, Katzman et al. 2018, Boyle et al. 2006, Murray et al. 2006, Malin-Mayor et al. 2022). While this signal could occupy a possibly redundant channel, it would bring extra phototoxicity. Accurate and effective tracing of the segmentation of cell shapes and lineages up to the 550-cell embryo (for *C.elegans*) from only fluorescently-labeled membranes would be desirable and could allow easier processing. Different contrast methods have been applied but do not allow manual editing in a reasonable time (Waliman et al. 2024, Sugawara et al. 2022), especially for late stage embryos.

Molecular Probes and Imaging Specifications for Cell Adhesion Quantification

For the quantitative analysis of cell adhesion, a dedicated transgenic *C. elegans* strain (ZZY0868) was engineered to enable simultaneous visualization of the plasma membrane and adherens junctions. The strain was constructed by crossing parental strains LP172 and ZZY0637, resulting in a genotype that incorporates two key fluorescent reporters: the core adherens junction protein HMR-1 was endogenously tagged with GFP (hmr-1(cp21[hmr-1::GFP + LoxP])), while the plasma membrane was specifically labeled with the PIP$_2$-binding pleckstrin homology domain of PLC1δ1 fused to mCherry (zzySi139 [Phis-72::PH(PLC1delta1)::mCherry::pie-1 3' UTR]). Embryos expressing both reporters were imaged live on a Leica SP5 laser-scanning confocal microscope, following the established miniMos-based method (Cao et al., *Nature Communications*, 2020). Sequential high-resolution z-stacks (XX slices per stack with a z-resolution of 0.42 μm) were acquired at intervals of approximately 1.5 minutes, using 488 nm and 561 nm laser lines for excitation of the HMR-1::GFP and PLC1δ1::mCherry signals, respectively. Z-axis compensation (19-95% for the 561 nm laser) was applied to mitigate signal attenuation. This imaging regimen captured the dynamic localization of adhesion complexes relative to cell morphology throughout early embryogenesis. The resultant four-dimensional data were exported as multi-TIFF stacks for subsequent processing and analysis by the MolQuantifier module.

**Detailed Training and Running Segmentation Workflow**

The running steps of *TUNETr*, with real nuclei fluorescence images, concludes recognizing membranes and generating nuclei-prompting seeds (Figure S1). The comparison of CTransformer is also conducted with the results under sTUNETr and nuclei-prompting strategy (Figures 2 and 3).

Step 1: Recognizing Cell Membrane with Transformer-based Regression Large Model and Topology-constraint Loss. In the initial phase of our method, as outlined in (step 1 of Figure S1A), we establish a ground truth (GT) dataset comprising manually annotated images, like other DNNs. These annotations represent a range of embryonic developmental stages and imaging conditions. This GT dataset serves as the foundation for training the *TUNETr* within *CTransformer*, resulting in a pretrained *iTUNETr* network, with 71 volumes in this study. We utilize the capability of this large-scale model, convoluting voxels of 3D volumes with their neighboring voxels (self-attention). So, *iTUNETr* facilitates the precise identification of voxels that were previously indistinguishable due to extremely weak signals, noises and various artifacts, as demonstrated in 5$^{th}$ columns of (Figure 3). Consequently, the recognition results in seen conditions (similar raw images in the GT) are granted by the DNN (*iTUNETr*). With pretrained *iTUNETr*, we introduce topology loss term, $L_t$ (step 1 in the right of Figure S1A), in the further training. Pretrained *iTUNETr* is necessary since that the topology loss can only be accurately computed with the context of a relatively correct spatial shape (cellular). This improved loss function is an enhancement of geometric loss term, $L_g$ (step 1 in the right of Figure S1A), being effective to constrain a single cell's topological structure in the same as the GT. This represents a minor adjustment to the initial model *iTUNETr*, from a higher-dimensional perspective (topological latent space). There are two ways to annotate the membrane for recognition. One approach starts from scratch, relying solely on human visual inspection to annotate images on each 2D slice and then observing in 3D to check if the annotations across different slices are continuous and credible. These annotations are voxel-wise accurate, but are subject to variability due to different annotators' standards and encounter issues with discontinuity in 3D. However, this can lead to issues where pre-recognized volumes influence human judgement, as some areas in the raw volumes are not absolutely clear, even with human annotation. So, we use both approaches to manually annotate the GT for training, to discard the dross and select the essence. Here, we get the "specialist" DNN, *iTUNETr*.

Steps 2 and 3: Synthetic Ground Truth of 4D Fluorescence Images and Fine-tuning Semi-supervised Learning Strategy. We trained the "specialist" DNN, using manually annotated GT, to recognize the membrane (semantic segmentation) for fluorescence 3D images. To overcome the biggest drawback of DNN, lacking generalization ability and only working at a special task, also for transformer based large models, we make a substantial synthetic GT and semi-supervised learning scheme that cooperates human intervention. The *iTUNETr* model recognizes cell membrane in anisotropic raw fluorescence images, achieving an accuracy comparable to human annotation level for seen imaging conditions (with Dice Score larger than 0.9). To amass a large GT dataset while maintaining a human level of accuracy, *TUNETr* allows users to manually adjust the nuclei promptings via StarryNite and AceTree(Boyle et al. 2006, Murray et al. 2008, Bao et al. 2006). This human-in-the-loop step provides location prompts and eliminates irrelevant surroundings from the recognition area, refining the membrane recognition to reasonable positions and sizes. Users can repeatedly tune the prompts and select the correct membrane recognition volumes to form the synthetic GT, integrating it with the manually annotated GT (first row of Figure 1b and step 2 of Figure S1a). These new GT membrane annotations are constraint from real cell shape, and the embryos' edges are tripled in thickness, for dragging DNN's attentions to the weak imaging area. This user-customized synthetic GT encompasses a vast of unlabeled data, covering all imaging conditions (both seen and unseen, labeled and unlabeled) relative to the task. Then, restoring from *iTUNETr*, adjusting the DNN parameters, and proceeding with training (step 3 of Figure S1a), *TUNETr* generates a "generalist" DNN, *sTUNETr* (semi-supervised *TUNETr*, akin to *iTUNETr*), tailored for specialized tasks under unseen imaging conditions, such as those with significant artifacts or in late developmental stages.

Steps 4: Generate Reliable Nuclei-prompting Seeds with Well-trained DNN. We introduce additional nuclei-prompting step to keep chain and serious cell shape segmentation errors away (step 4 of Figure S1a). Informed by nucleus promptings, *TUNETr* avoided completely erroneous segmentation in extremely ambiguous imaging areas. This step helps in constructing potent cell instance generation (multi-cell segmentation) algorithm from cell membrane recognition. Due to the DNN model is an image-wise global and customized method, the recognition of previous studies for top and bottom areas (axially upper and lower) in the volumetric images still failed when the imaging conditions is very different or the trained model has not "met" before(Eschweiler, Smith and Stegmaier 2022, Wang et al. 2022, Cao et al. 2020) (in the training data). Fortunately, nucleus images are relatively clear and easy to detect as nucleus instance. Thus, we significantly improve the robustness of the mentioned DNNs (Table S11), with or without nuclei-promptings.

Steps 5 to 7: Producing Multiple Single-cell Segmentation for 4D Fluorescence Images. At the final processing steps (steps 5 to 7 of Fig. 1a), we run the segmentation on 34 embryos, providing a dataset for fluorescence images processing and relative biomedical studies. Integrating the large training set and *sTUNETr*, we generate a more convincing cell morphology map. In this work, we provide interface software and corresponding website for this 4D cellular segmentation map, also compatible with the visualization database website (CMOS bcc.ee.cityu.edu.hk/cmos) and analysis software (ITK-SNAP-CVE)(Cao et al. 2020, Guan et al. 2023). The running multiple cell segmentation program, the interface software is relatively friendly to non-programmers, biologists and physicists. Therefore, for different multicellular worm *C. elegans*, *TUNETr* system can directly generate 4D data that can be traced back to the cell shape level, which can greatly promote the study of cell morphology in the embryonic development process.

Conclusively, under different imaging conditions, for the ambiguous situation and late developmental time point, the images suffered from unavoidable laser attenuation and cells have grown in deformed shape, which is far from the images and shape in the training data. We demonstrate the leading results of membrane recognition, cellular segmentation, cell shape generation, and reconstruction capability for two most challenging cell fates (Figure 2 and 3). We also study the robustness of nuclei-prompting module (Table S11), with raw fluorescence nuclei images. Even with fuzzy input 3D volumes or unsuitable cell membrane recognition methods, this module keeps the results from degrading too quickly. During a long-range developmental period, from 4— to 550—cell embryogenesis, *TUNETr* shows the ability to segment difficult cell morphology with good robustness, which also validates the generalizability of our method.

## Topological Wasserstein Distance Loss

Transformer (attention module) deep neural network solves the local features learning problem. However, the existing loss functions in voxel-wise recognition (segmentation), like the binary cross entropy (BCE), mean square error (MSE), and soft Dice score loss in VNet(Milletari, Navab and Ahmadi 2016), are not coherent to regularizing the cellular boundary and have no receptive field in global 3D images. To drive the network to not only geometrically accuracy but also cellular shape level correct, we propose a topology-constraint loss integrating EDT, persistent homology (PH), and $p$ —Wasserstein distance(Panaretos and Zemel 2019). The proposed loss includes 2 parts: geometrical $L_g$ and topological loss $L_t$:

$$L_\epsilon(Y_{pred}, Y_{gt}) = L_g(Y_{pred}, Y_{gt}) + w_\epsilon \times L_t(Y_{pred}, Y_{gt}) + \xi,$$

where the $\epsilon$ representing the current training epochs, $Y_{pred}$ and $Y_{gt} \in R^{(H \times W \times D)}$ are the predicted and ground truth cell membrane volumes respectively, and $W_\epsilon$ is the weight of topological loss according to the training epochs. $\xi$ is the necessary constant term for all loss functions. The geometrical loss is calculated by MSE: $L_g(y_{pred}, y_{gt}) = \sum |\mu_{(h',w',d')} - v_{(h',w',d')}|$, where $\mu_{(h',w',d')}$ and $v_{(h',w',d')}$ are respectively the probabilities of every voxel in $y_{pred}$ and $y_{gt}$. Topological information and features can be extracted and learned by the *TUNETr* when the prediction is closed to the ground truth, so the geometrical loss is the main loss in $L_\epsilon$ at the initial training part. $w_\epsilon$ is a consistent weight for topological loss depending on the complexity of the images (the number of topological structures).

$L_t$ from cubical persistent homology (PH) diagram $D$, is calculated to extract topological structural difference of the prediction and ground truth 3D images (Euclidean distance transformed). For the volumetric images, they would be transformed to cubical complexes for representing images in a set of vertices, edges, facets and other higher−dimensional counterparts, which are basic topological elements. PH, is counting the number of $p$−dimensional topology structures ($\beta_k$), also known as $k^{th}$ Betti numbers in $D_{pred}^{(k)}$ and $D_{gt}^{(k)}$, including the birth and death time of $\beta_k$ (ref the PH showing figures), in the probability map. In the 3D space, there are 3−dimensional topological features, $\beta_0, \beta_1, \beta_2$ are the number of connected components (1D), holes (2D) and hollow voids (3D, the similar 3D structure in torus). The PH diagram is defined as $D = \sum_{k=0}^{p} D^{(k)}$ of an image.

Wasserstein distance is *optimal transport* or the *Monge − Kantorovich* problem assessing the path, "difference" and distance from persistent homology diagram $D_{pred}$ of $Y_{pred}$ to $D_{gt}$ of $Y_{gt}$ [CITE PYTORCH TOPOLOGY PERSISTENT PACKAGE]. $L_t$ is derived from $p$−Wasserstein distance between probability maps ($p = 3$ for 3D images) with stability to noise(Waibel et al. 2022, Cohen-Steiner et al. 2010, Panaretos and Zemel 2019):

$$W_p\left(D_{pred}^{(k)}, D_{gt}^{(k)}\right) = \left(inf_{\eta: D_{gt}^{(k)} \to D_{pred}^{(k)}} \sum_{x \in D_{gt}^{(k)}} ||x - \eta(x)||\right)^{1/p},$$

where $\eta(x)$ denotes a bijection from $D_{gt}$ to $D_{pred}$. $D_{gt}$ and $D_{pred}$ have different $\beta_k$, the bijection denotes to that the points of $D_{pred}$ would match with the closest point in $D_{gt}$. If the extra points have no match in $D_{gt}$, these points would be compared with diagonal. Therefore, for $D_{pred}$ of the prediction $Y_{pred}$ and $D_{gt}$ of the manually annotated ground truth $Y_{gt}$, the topology-constraint loss is formulated as :

$$L_t = \sum_k^p W_p\left(D_{pred}^{(k)}, D_{gt}^{(k)}\right).$$

Geometrical loss term $L_g$ penalizes globally voxel-wise mismatches between *TUNETr*'s predictions and the GT and topological term $L_t$ encourages the predicted probability map to generate the similar amounts and structures of topology persistent homology with the GT.

**Semi-supervised Learning with Edge-enhancing Synthetic GT for Boundary-aware Model**

To address these significant obstacles and failure of recognizing boundary membrane, insufficient annotated GT, weak generalization and low accuracy for supervised learning in processing 3D + *t* fluorescence images, we integrate semi-supervised learning and utilize the large number of unannotated images in *TUNETr*. We trained *iTUNETr* from a few manual GT with nuclei prior knowledge and *iTUNETr* synthesize manually-annotated level synthetic GT from the massive remaining 3D images. From the multi-cell segmentation from *iTUNETr*, the clear binary cell membrane of time-lapse volumes were constructed with shape-constraint module to select human-annotated level GT for *sTUNETr* (synthetically semi-supervised learning model, a part of *TUNETr*). This module would filter the volumes whose predicted probability maps are far away from their real topological structures and triple the fluorescence signal intensity in the outermost membrane of the GT to combat serious signal weakening problems. With *TUNETr* boosts the potential and performance of our model with only 71 labelled images. Even at the late-stage imaging domain, *TUNETr* provides human-level membrane recognition and cellular segmentation (Figures 2 to 4). As illustrated in the main text, for 16 evaluation GT, *CTransformer* achieves 4.1% cell loss rate. And the improved parts are mainly located at the skin and apoptotic tissues. Moreover, considering some manually annotated errors, the Dice Score 0.9 of *CTransformer* has reached the human-level accuracy. This training strategy nuance handling of spatial information and its sensitivity to cellular boundaries ensure that the integrity of cellular morphology is meticulously preserved, thereby enhancing the overall accuracy and reliability of the segmentation process.

**Transformer-based Large Model**

*TUNETr*, a UNet-like Swin Transformer deep learning neural network, is inspired by previous work (Figure S1b), UNet, Transformer, Swin Transformer, and SwinUNETR(Ronneberger et al. 2015a, Liu et al. 2021, Hatamizadeh et al. 2022a), for 3D + t cellular membrane recognition. Contracting images to high-level spatially relative feature maps, *TUNETr* consists of a stack of transformers as the encoder. Embedding 1D sequence input from a volumetric image, *TUNETr* partitions a 3D image $x \in R^{(H \times W \times D)}$ into $N = \frac{H \times W \times D}{h \times w \times d}$ one-dimensional (1D) sequences, $x_v \in R^{(h \times w \times d \times C)}$, also named patches or tokens, which is representing a word in transformer of NLP. Here, $(H, W, D)$ and $(h, w, d)$ denote the resolutions of the 3D image and each patch, and $C$ is the number of embedding spaces of feature map (feature size). Self-attention computation is conducted in non-overlapped $(h \times w \times d \times C)$ patches (tokens) and the small-sized patches for reasonable computational complexity but it has no communication with other patches while losing global attention. To obtain significant global attention in density prediction, *TUNETr* applied the shifted window based self-attention in the volumetric semantic segmentation. One window (layer $l$) contains $(M, M, M)$ patches evenly in $N$ patches from a 3D image, and the next layer $l+1$, is shifted with $(\frac{M}{2}, \frac{M}{2}, \frac{M}{2})$ voxels. The previous and subsequent shifted windows, layer $l$ and $l+1$, both containing $M \times M \times M$ patches, with window based multi-head self-attention (*W-MSA*) and shifted *W-MSA* (*SW-MSA*)(Liu et al. 2021, Hatamizadeh et al. 2022a) is calculated by

$$\hat{z}^l = W\text{-}MSA\left(LN(z^{l-1})\right) + z^{l-1},$$

$$z^l = MLP\left(LN(\hat{z}^l)\right) + \hat{z}^l,$$

$$\hat{z}^{l+1} = SW\text{-}MSA\left(LN(z^l)\right) + z^l,$$

$$z^{l+1} = MLP\left(LN(\hat{z}^{l+1})\right) + \hat{z}^{l+1},$$

where the $\hat{z}^l$ and $\hat{z}^{l+1}$ denote the outputs of *W-MSA* and *SW-MSA* modules, providing cross-window (global attention of one volume) connections. Multilayer perceptron module (*MLP*) convolutes the data to a target resolution (number of channels) and linear normalization module (*LN*) avoids the vanishing and exploring gradient problems. *TUNETr* also utilizes cyclic and reverse cyclic shift in masking and padding the hierarchical window partitions [Cite Swin transformer] for efficient self-attention calculating without messing up the original image order. To keep the significant spatial inductive bias, self-attention is calculated by $Attention(Q, K, V) = Softmax(QK^T/\sqrt{b})V$ where $Q, K, V \in$

$R^{(h\times w\times d\times b)}$ are the query, key and value metrices in transformer, and $b$ is the query/key resolution(Liu et al. 2021, Hatamizadeh et al. 2022a).

The encoder of *TUNETr* sets $(h, w, d)$ of each patch as (2 x 2 x 2) and a patch length is $2 \times 2 \times 2 \times 2 = 16$, with 2 nucleus and membrane channels of the 3D volumetric images. With raw input 3D image $x \in R^{(128\times 128\times 128)}$, the feature size of embedding space $C$ as 48, and depths of 4 stages as (2, 2, 2, 2). Swin Transformer Blocks (respectively 8 *W-MSA* and *SW-MSA* modules)(Liu et al. 2021), *TUNETr* encoder processes the input $x \in R^{(128\times 128\times 128\times 2)}$ as following stages (Swin transformer block): (64, 64, 64, 48), (32, 32, 32, 96), (16, 16, 16, 192), (8, 8, 8, 384), (4, 4, 4, 768). Here, keeping explainable hierarchical and consistent with UNet, a patch merging module reduces the spatial resolution by 2 while the channel number would be embedded and increased by 4, and a linear convolutional neural layer reduces channel number by 2 for output of each stage (Swin Transformer Block).

The corresponding decoder integrating residual layers, increasing the resolution of feature maps, and receiving skip connections from stages of encoder, the upper data flow from high level to low level semantic (understandable) feature map denotes as: (8, 8, 8, 384), (16, 16, 16, 192), (32, 32, 32, 96), (64, 64, 64, 48), (128, 128, 128, 48). The output feature map at the bottleneck is fed into a convolutional layer and a Euclidean distance transform global regression module (EDT-GFR) for cellular membrane recognition (prediction of probabilities of voxels being cell membrane).

**Generating Multi-nuclei Segmentation GT and Training the Model**

We meticulously extracted embryonic nuclei tracing data from high-resolution 4D Fluorescence Imaging, as delineated in Section 4D Fluorescence Images and Cell Identities Acquisition of main text. This process entailed the precise determination of nuclear centroids, identities, and volumetric parameters. For the annotation of the nuclei ground truth (GT) dataset, encompassing 1270 *in vivo* embryonic volumes, we employed a multi-nucleus segmentation protocol. Each nucleus was annotated in three-dimensional space using spherical representations, meticulously ensuring consistency in resolution (H×W×D) with the corresponding cell membrane images.

Confronting the challenge of nuclear overlap, particularly pronounced in datasets beyond the 150-time point, we innovatively adapted our methodology. This adaptation involved halving the radii of the nuclear spheres, a strategic modification that significantly enhanced the spatial resolution of individual nuclei within densely populated regions.

Further, for the processing of nuclei GT, we leveraged the star-convex polygon model, as conceptualized in StarDist(Schmidt et al. 2018), extending it to encompass 96 radial vectors. This approach facilitated the generation of a nuanced gradient map, capturing topological variances and the radial distances from each voxel to the nearest nuclear centroid. Integrating the advanced StarDist3D framework(Weigert et al. 2020), into our *TUNETr* system was pivotal (Figure S15). This integration was tailored for the nucleus-prompting mechanism, utilizing star-convex polygons with 96 radial vectors to render a topologically informed gradient map. This map was pivotal in conveying spatial relationships, delineating equidistant angular separations, and quantifying voxel-centric nuclear proximities.

For the optimal training of the star-convex polygon model, we employed a hybrid loss function, synergistically combining Mean Absolute Error (MAE) and Binary Cross-Entropy (BCE). This dual-functional loss function was instrumental in refining voxel classification and probability prediction. To augment the fidelity of our model's output, we implemented an advanced Non-Maximum Suppression (NMS) algorithm. This algorithm was critical in curtailing redundancy within the predicted volumetric data, thereby ensuring the integrity and precision of our segmentation outcomes.

**Running Dual-approach Strategy with Nucleus-prompting**

| **Algorithm 1: Delaunay-clustering algorithm.** |
|---|
| Require: $I_{pred}$, $h_{thres}$ |
|     get locally minima points $P_{min}$ with $h_{thres}$ |
|     get whole-cell segmentation set $S(cell)$ |
|     $S(cell) = watershed(I_{pred}, P_{min})$ |
|     build graphical edges set $S(edges)$ |
|     $S(edges) = Delaunay(S(cell))$ |
|     **while** $edge$ in $S(edges)$ **do** |
|         sum all overlapped and connected components |
|         calculate set of edge weights $\omega_{edge}$ |
|     **end while** |
|     drop and merge the overlapped meshes |
|     return regular cells in $S(cell)$ |

To conquer the challenging multicell object segmentation from the occasionally incorrect recognition of cell membrane signals, *TUNETr* utilizes the 4D multi-nuclei segmentation from StarDist3D(Weigert et al. 2020) and Delaunay-clustering watershed algorithm (Figure S2 and Algorithm 1) in 4D membrane recognition (semantic segmentation) to generate robust and accurate multiple whole-cell regions (multi-cell segmentation).

Using independent location seeds unlike previous work(Cao et al. 2024, Cao et al. 2020, Guan et al. 2023), we utilized the Delaunay clean method from them, where pixel clusters are organized into graphs using Delaunay triangulation. Without binarizing the grayscale probability maps and finding explicit nuclei position as local minima, which is suffering from implicit membrane recognition results, this step of *TUNETr* groups points near cell margins and assigns weights to graph edges. Overlapping cells (crossed edges and vertices) are also filtered out to achieve a cleaner segmented map. Conclude with marker-watershed segmentation using the refined seeds.

*TUNETr* capitalizes on the advanced capabilities of the marker-seeded watershed algorithm(Li et al. 2010, Van Der Walt et al. 2014), to integrate experimentally-determined multi-nucleus segmentation(Schmidt et al. 2018), into its computational framework. This integration is manifested in the refinement of the predicted probability map, hereafter referred to as $Y_{pred}$. The incorporation of multi-nucleus segmentation serves a dual purpose: firstly, it provides precise, automated localization and quantification of cellular dimensions, and secondly, it acts as a critical navigational aid for the segmentation of multiple cellular objects within complex morphological structures.

## Method Implementation Details and Evaluation Experimental Setup

*TUNETr* is partly implemented with PyTorch(*Paszke* et al. 2019), StarDist3D(Weigert et al. 2020), SwinUNETR(Hatamizadeh et al. 2022a) and MONAI(Cardoso et al. 2022). m2nGAN is partly built on 3D Cycle GAN(Ge et al. 2019, Isola et al. 2017).

Both *TUNETr* and *m2nGAN* are trained on a Dell computation station with 2 NVIDIA RTX A6000 GPUs (48GiB), 2 Intel Xeon Gold 6248R CPUs (3.00GHz), and 1TB physical memory. GPU driver version is 470.199.02 and CUDA version is 11.4. Python is 3.9.16, with 1.10.1+cu111 torch, 0.10.1+cu111 torchvision and 1.2.0rc7+22.g575954cb *MONAI*. The 3D *iTUNETr*, *sTUNETr*, and the StarDist3D model (all the trained DNNs model) in the end-to-end method *TUNETr* (the proposed segmentation method), as well as other comparing SOTA DNN models, are trained with the same computational resource and augmentation operation. The membrane recognition models are trained with 5000 epochs, $160 \times 160 \times 160$ randomly cut volumes inputs. By perturbing the pixels' intensity (uniform distribution, scaled with the half-open interval $[1,1.1)$ and shifted with $[0,0.1)$) and randomly flipping (50% chance to be flipped along the *x*, *y*, and *z* axes, respectively), the 71 3D manually annotated volumes in the training dataset were augmented to multiple different and effective training 160×160×160 cube images. The Adam optimizer is conducted to update the network with an initial learning rate of $5 \times 10^{-3}$ and a weight decay rate of $1 \times 10^{-5}$, using AMSGrad gradient descent optimization.

The evaluation experiments are conducted on 16 manually annotated volumes. The cell morphology of five embryos "WT_C_Sample2", "WT_C_Sample3", "WT_C_Sample4", "WT_Sample1",and "WT_Sample7" was annotated by seven well-trained experts, from *CShaper*(Cao et al. 2020), *CMap*(Guan et al. 2023) and the raw fluorescence images. They were re-grouped and organized. The segmentation results were manually checked and corrected slice by slice (2D view) and cell by cell (3D view), ensuring the correctness and smoothness of 3D cell shapes in the evaluation GT. Specifically, we gained new time lapse and cell-number-wise GT data. First, the middle slice at each imaging time point throughout embryogenesis (255 2D images for "WT_Sample1" and 205 for "WT_Sample7") was annotated for 2D comparison, providing the cross-section of 30509 cell regions in total. Second, the complete typical 3D volume within 100±5−, 200±5−, 300±5−, 400±5−, 500±5−, and 550±5−cell stages (2 late-stage 3D images for either "WT_Sample1" or "WT_Sample7", 3 early stage for "WT_C_Sample2", "WT_C_Sample3", or "WT_C_Sample4") were annotated for 3D comparison, providing the full morphology of 4046 cell regions in total.

## *CTransformer* Effectively Segmented Elongated and Engulfing Cells

Achieving machine-based predictions with accuracy similar to human-level annotations in 550-cell embryos remains a challenge. During embryogenesis, cells can undergo significant morphological changes, such as elongating and aligning in grid patterns, with deformations, particularly from spherical to elongated shapes, posing considerable challenges for standard DNN and cell instance generation algorithms. Skin cells form elongated shapes and move to the outer edge of the embryo and cover other cells from the 350-cell stage. After the 500-cell stage, skin cells cover more than 60% of the embryo's surface and continue to grow and close. Apoptotic cells shrink and will be engulfed by a surrounding cell and be embedded in other cells giving highly dense membrane signals, which result in unusual and ambiguous areas that are difficult to generate. Tissues, like "Other" and "Neuron", also proliferate at small cell sizes. *CTransformer* is able to manage irregular shapes and nearly invisible regions.

These time-lapse fluorescence images, affixed to a glass slide with polylysine, were generated from the ZZY0637 and ZZY0655 strains on NGM (Nematode Growth Media) plates. The embryos were seeded with OP50 and kept at room temperature, 20◦C (Table S1). During imaging, the temporal resolution was 1.39 to 1.45 min per frame, and the volumetric spatial resolution was 0.18 or 0.25 ($0.18^3$ or $0.25^3$ $\mu m^3$ per voxel).

Practical and efficient scalable software is provided for time-lapse single-cell volumetric analysis, and 4D embryonic segmentation from fluorescent images of embryos. The code, software, raw and segmented data are available in link https://doi.org/10.6084/m9.figshare.27085657.

# Additional Review and Results

Deep Learning-based Related Works

Manual annotation of 3D volumetric images is labor-intensive and time-consuming making obtaining sufficient large-scale time-lapse cell morphological data for training impractical(Cao et al. 2024, Cao et al. 2020, Guan et al. 2023, Liu et al. 2021) and suitable training dataset is not available. Late stage developmental embryos contain many cells with deformed shapes and anisotropic sizes, further complicating building a single-cell morphology map. At late stages, a slight recognition error of a cell membrane might cause a large area of cells to be incorrectly connected or lost, with possible exponential amplification of the error(Cao et al. 2019, Cao et al. 2024, Cao et al. 2020). In low-resolution cellular images, existing DNN methods, the data-driven models, could be improved to recognize the membrane signals in data with different distribution.

However, as cells divide, nuclei are positioned more closely together in cells with small sizes that can give unbalanced, ambiguous signals. Frequent and prolonged imaging is necessary to trace cells, which leads to photobleaching, the loss of fluorescent intensity over time, and phototoxicity that might damage cells and disrupt development. In *C. elegans*, the expression of a fused histone and marker protein can be too weak to detect in sublineages, like the progenies of D and E cells. Stronger expression might kill the embryo and reducing laser intensity might not give sufficient signal. Thus, the balance between signal quality and embryonic health influences the ability to establish and quantitatively analyze a map of the patterns of cellular morphology and gene expression .

Through balancing tuned images, advanced optical and deep learning segmentation algorithms have been established[4-7], however, their ability to recognize cell membrane signals and reconstruct morphological maps still has limitations. Under normal circumstances, the maps are derived from images of fluorescently labeled membranes and nuclei that give cell positions and detailed cellular shapes. However, occasional laser attenuation during imaging, insufficient 3D manually annotated ground truth (GT) data to develop a deep learning model, and the inability of a deep neural network (DNN) to recognize weak local signals, remain problems.

Fully supervised convolutional and transformer based DNNs(Ronneberger et al. 2015b, Milletari et al. 2016, Chen et al. 2019, Zhou et al. 2020, Isensee et al. 2021, Hatamizadeh et al. 2022a, Hatamizadeh et al. 2022b, He et al. 2023), were proposed for pixel or voxel recognition (semantic segmentation), while the literature has not extensively tackled the above-mentioned recognition and segmentation challenges. The first 5 DNNs(Ronneberger et al. 2015b, Milletari et al. 2016, Chen et al. 2019, Zhou et al. 2020, Isensee et al. 2021), U−Net, VNet, 3DUNet++, DMFNet, and nnU-Net, are based on convolutional neural network (CNN) structure, which encoded feature maps with spatially multi-scale communication and locality inductive bias, requesting a few ground truth (GT) for biomedical imaging analysis. The U-shaped DNNs expanded the depth of the neural network. The latter 3 DNNs(Hatamizadeh et al. 2022a, Hatamizadeh et al. 2022b, He et al. 2023), UNETR, SwinUNETR, and SwinUNETR−V2, are based on transformer and swin-transformer(Vaswani et al. 2017, Liu et al. 2021), create a local content-aware attention mechanism to leverage the context of the input patches (words or tokens in natural language processing, NLP). The breadth of the neural network has been enlarged by transformer. Then for multiple cell objects segmentation, competitive methods were developed(Stegmaier et al. 2016, Azuma and Onami 2017, Cao et al. 2019, Cao et al. 2020), involving training one or a collection of specialized deep neural network (DNN), each tailored to a distinct imaging condition or animal as "specialist" approach. "Specialist" DNN could not be extended well and directly to other imaging situation. Some "generalist" DNN approaches are proposed for general single cell segmentation(Schmidt et al. 2018, Weigert et al. 2020, Stringer et al. 2021, Eschweiler et al. 2022, Wang et al. 2022), to obtain multi-cell segmentation. They employ semantic segmentation of the cell membrane and instance segmentation as post-processing. StarDist and Cellpose are designed and adapted for 2D fluorescence microscopies(Schmidt et al. 2018, Stringer et al. 2021). StarDist3D and Cellpose3D are applied to 3D, transforming the image domain and generates gradient feature maps to create 3D cell instances(Eschweiler et al. 2022, Stringer et al. 2021), and assisted by star-convex and polygon theories, detecting the voxel-wise outline of a cell and generates fixed instances using an interpretable algorithm(Weigert et al. 2020, Schmidt et al. 2018). Unfortunately, these methods are still unable to recognize extremely fuzzy membranes and establish irregular cell morphology in low-resolution images of living embryos. These DNNs address the problem of fluorescence signals recognition in blurred and unrecognizable microscopic images. However, all of above methods rely heavily on large amounts of annotated data; they can only play a role of identification and preliminary (semantic) segmentation for good-quality images, lacking the capability of pushing method to noisy and ambiguous images. Moreover, they have not been fitted generally or the models are too small to handle complicated situations.

Solving Intrinsic Fluorescence Heterogeneity and Axially Weak Boundary Signals Membrane Recognition

First, when we conduct fluorescence imaging and optical scanning 3D embryos, there were some intrinsic light heterogeneity and unbalanced weak signals. The fluorescence heterogeneity was shown in the first column of (Figure S6). The brightness of cell membrane signals in the inner and middle layer were very different, because the inner cells are larger and therefore their signaling proteins (mCherry here) expressions were less densely dispersed. This intrinsic anisotropy would reduce a big inner cellular segmentation failure and severe chain cell loss. Second, for living embryos, laser beam was always tuned down for ordinarily embryonic development, which would cause weak and ambiguous signals somewhere. The deficiencies in these images can be particularly severe in some areas (Figure S6, c and d, >550−cell stage volumes). Meanwhile, restricted by the 3D scanning confocal imaging (2D slice by slice), axial direction (perpendicular to the shooting plane) was in low resolution. It was difficult to maintain spatial consistency of a living embryonic volume. Additionally, the areas circled by the large white rectangular box were the areas where the cell signal is severely blurred or disappeared, mainly in the last few layers of the slice for pseudo-3D imaging (Figure S6).

We proposed the topological loss function, cooperating with the large-scale model and semi-supervised and edge-enhancing learning for the cellular segmentation of 4D live-cell embryos (see Section Transformer-based Large Model, Topological Wasserstein Distance Loss, and Semi-supervised Learning with Edge-enhancing Pseudo-GT), to solve these imbalances by producing membrane recognition with consistent recognized results (the *iTUNETr* column of Fig. 3). Compared with images of other organisms, the membrane recognitions were in much lower anisotropic brightness distribution, no matter axially or sagittally. The signal strengths recognized by the inner, middle, and outer cell membranes of the embryo are relatively balanced. Moreover, the membrane recognition of *CTransformer* (*sTUNETr* column in Figure 3), presents its comprehensive 3D membrane recognition for laser attenuated embryonic volumes.

For the first point, low intensity and contrast signals also exist in the outermost layer of the cell membrane. This is due to two reasons. The first is because the inner cell membranes are in contact with each other, and there are two cell membranes with fluorescent protein expression; whereas the outermost cell membrane is in contact with the outside environment, which is only one layer of cell membrane, and the protein expression is only half of that on the inside, nearly disappeared (1$^{st}$ column of Figure 3b). This makes it very confusing for the DNN model to know if there are membranes. The second reason is that when the optical equipment scans the image layer by layer axially, there is a serious light scattering problem in the start and end layers of the scanning (1$^{st}$ column of Figure 3b). This is also a problem that image restoration and super resolution in post-processing are trying to solve, but they are still not conquered (for general images). *TUNETr* is effective in solving this weak boundary signals problem that the outermost cell membrane is difficult to recognize. Thanks to the transformer-based and large-scale model, our model can pay higher attention to thicken the membrane contacting with the environment, as shown in the 6$^{th}$ column of Figure 3. Additionally, with the above-mentioned topological loss, *iTUNETr* connects the incomplete and invisible membrane as well, by learning the cell membrane skeleton of the whole embryo in the topological latent space, which is relatively intact, rather than fragmented as in other models.

Furthermore, for second point we are presenting the capability of *sTUNETr* for generating invisible membrane and axially undersampling signals. Qualitatively, *sTUNETr* allows extract foreground (membrane) explicitly, without uncertain prediction around the membrane skeleton. For the axially (z direction) low-resolution issue (Figure 3c), semi-supervised learning produced a lot of pseudo-GT data with topological priors (see Section Generating Multi-nucleus Segmentation GT and Training the Model), to provide detailed 3D cellular morphological information axially. With a few annotated images and sufficient pseudo training data, *sTUNETr* reconstructed the 3D cell membranes in ambiguous images. At late stage time points, cells become extremely small, containing only less than 10 voxels or 3 axial pixels, in some cases. Thus, *sTUNETr* improving axially undersampling signals significantly enhances the constructions of not only ordinary cells but also irregular cell shapes, which are everywhere in large numbers in late-stage embryos. For laser attenuated areas, they can be successfully recognized and segmented (Figure S6).

**Ablation Analyses**
    a) The basic deep learning model is the transformer-based regression network (base TUNETr).
    b) topological loss (iTUNETr – trained model called iTUNETr at the previous subsection),
    c) semi-supervised shape-constraint (EDT-Swin-UNETR + TL + SC-SL – trained model called sTUNETr in other sub-section)

d) nucleus-prompting seeds (EDT-Swin-UNETR + TL + SC-SL + NPS – the last TUNETr)

1 Topological loss improves 5% membrane segmentation accuracy
2 Semi-supervised make the approach balanced distributed and robust in mathematics
3 Nuclei segmentation help a lot on generating multiple cell objects from cell membrane semantic segmentation

### *CTransformer* Interface/software

In this work, we provide a software interface for researchers to run the segmentation, cell shape analysis and visualization without coding or configuring GPUs (Figure S14). Because many biologists or researchers know little or have few time to coding, *CTransformerApp* enable feasible tools for them to use our method.

# References

Azuma, Y. & S. Onami (2017) Biologically constrained optimization based cell membrane segmentation in *C. elegans* embryos. *BMC Bioinformatics,* 18.

Bao, Z., J. I. Murray, T. Boyle, S. L. Ooi, M. J. Sandel & R. H. Waterston (2006) Automated cell lineage tracing in Caenorhabditis elegans. *Proceedings of the National Academy of Sciences,* 103, 2707-2712.

Boyle, T. J., Z. Bao, J. I. Murray, C. L. Araya, R. H. Waterston, T. J. Boyle, Z. Bao, J. I. Murray, C. L. Araya & R. H. Waterston (2006) AceTree: a tool for visual analysis of Caenorhabditis elegans embryogenesis. *BMC Bioinformatics 2006 7:1,* 7.

Cao, J., G. Guan, V. W. S. Ho, M.-K. Wong, L.-Y. Chan, C. Tang, Z. Zhao & H. Yan (2020) Establishment of a morphological atlas of the *Caenorhabditis elegans* embryo using deep-learning-based 4D segmentation. *Nature Communications,* 11.

Cao, J., L. Hu, G. Guan, Z. Li, Z. Zhao, C. Tang & H. Yan (2024) CShaperApp: Segmenting and analyzing cellular morphologies of the developing Caenorhabditis elegans embryo. *Quantitative Biology*.

Cao, J., M.-K. Wong, Z. Zhao & H. Yan (2019) 3DMMS: robust 3D Membrane Morphological Segmentation of C. elegans embryo. *BMC Bioinformatics,* 20.

Cardoso, M. J., W. Li, R. Brown, N. Ma, E. Kerfoot, Y. Wang, B. Murrey, A. Myronenko, C. Zhao & D. Yang (2022) MONAI: An open-source framework for deep learning in healthcare. *arXiv preprint*.

Chen, C., X. Liu, M. Ding, J. Zheng & J. Li. 2019. 3D Dilated Multi-fiber Network for Real-Time Brain Tumor Segmentation in MRI. In *Lecture Notes in Computer Science*, 184-192. Springer International Publishing.

Cohen-Steiner, D., H. Edelsbrunner, J. Harer & Y. Mileyko (2010) Lipschitz Functions Have L p -Stable Persistence. *Foundations of Computational Mathematics,* 10, 127-139.

Eschweiler, D., R. S. Smith & J. Stegmaier. 2022. Robust 3d Cell Segmentation: Extending The View Of Cellpose. In *2022 IEEE International Conference on Image Processing (ICIP)*. IEEE.

Fickentscher, R., P. Struntz & M. Weiss (2016) Setting the Clock for Fail-Safe Early Embryogenesis. *Physical Review Letters,* 117.

Fire, A. (1994) A four-dimensional digital image archiving system for cell lineage tracing and retrospective embryology. *Bioinformatics,* 10, 443-447.

Ge, Y., D. Wei, Z. Xue, Q. Wang, X. Zhou, Y. Zhan & S. Liao. 2019. Unpaired Mr to CT Synthesis with Explicit Structural Constrained Adversarial Learning. In *2019 IEEE 16th International Symposium on Biomedical Imaging (ISBI 2019)*. IEEE.

Guan, G., Z. Li, Y. Ma, J. Cao, M.-K. Wong, L.-Y. Chan, H. Yan, C. Tang & Z. Zhao. 2023. Quantitative cell morphology in *C. elegans* embryos reveals regulations of cell volume asymmetry. Cold Spring Harbor Laboratory.

Guan, G., M.-K. Wong, Z. Zhao, L.-H. Tang & C. Tang (2021) Volume segregation programming in a nematode's early embryogenesis. *Physical Review E,* 104, 054409.

Hatamizadeh, A., V. Nath, Y. Tang, D. Yang, H. R. Roth & D. Xu. 2022a. Swin UNETR: Swin Transformers for Semantic Segmentation of Brain Tumors in MRI Images. In *Brainlesion: Glioma, Multiple Sclerosis, Stroke and Traumatic Brain Injuries*, 272-284. Springer International Publishing.

Hatamizadeh, A., Y. Tang, V. Nath, D. Yang, A. Myronenko, B. Landman, H. R. Roth & D. Xu. 2022b. UNETR: Transformers for 3D medical image segmentation. In *Proceedings of the IEEE/CVF winter conference on applications of computer vision*, 574-584.

He, Y., V. Nath, D. Yang, Y. Tang, A. Myronenko & D. Xu. 2023. SwinUNETR-V2: Stronger Swin Transformers with Stagewise Convolutions for 3D Medical Image Segmentation. In *Lecture Notes in Computer Science*, 416-426. Springer Nature Switzerland.

Isensee, F., P. F. Jaeger, S. A. A. Kohl, J. Petersen & K. H. Maier-Hein (2021) nnU-Net: a self-configuring method for deep learning-based biomedical image segmentation. *Nature Methods,* 18, 203-211.

Isola, P., J.-Y. Zhu, T. Zhou & A. A. Efros. 2017. Image-to-image translation with conditional adversarial networks. In *IEEE Conference on Computer Vision and Pattern Recognition (CVPR)*, 1125-1134.

Kretzschmar, K. & F. M. Watt (2012) Lineage Tracing. *Cell,* 148, 33-45.

Li, D., G. Zhang, Z. Wu & L. Yi (2010) An Edge Embedded Marker-Based Watershed Algorithm for High Spatial Resolution Remote Sensing Image Segmentation. *IEEE Transactions on Image Processing,* 19, 2781-2787.

Liu, Z., Y. Lin, Y. Cao, H. Hu, Y. Wei, Z. Zhang, S. Lin & B. Guo. 2021. Swin transformer: Hierarchical vision transformer using shifted windows. In *Proceedings of the IEEE/CVF international conference on computer vision*, 10012-10022.

Ma, X., Z. Zhao, L. Xiao, W. Xu, Y. Kou, Y. Zhang, G. Wu, Y. Wang, Z. Du, X. Ma, Z. Zhao, L. Xiao, W. Xu, Y. Kou, Y. Zhang, G. Wu, Y. Wang & Z. Du (2021) A 4D single-cell protein atlas of transcription factors delineates spatiotemporal patterning during embryogenesis. *Nature Methods 2021 18:8,* 18.

Malin-Mayor, C., P. Hirsch, L. Guignard, K. McDole, Y. Wan, W. C. Lemon, D. Kainmueller, P. J. Keller, S. Preibisch, J. Funke, C. Malin-Mayor, P. Hirsch, L. Guignard, K. McDole, Y. Wan, W. C. Lemon, D. Kainmueller, P. J. Keller, S. Preibisch & J. Funke (2022) Automated reconstruction of whole-embryo cell lineages by learning from sparse annotations. *Nature Biotechnology 2022 41:1,* 41.

Mcdole, K., L. Guignard, F. Amat, A. Berger, G. Malandain, L. A. Royer, S. C. Turaga, K. Branson & P. J. Keller (2018) In Toto Imaging and Reconstruction of Post-Implantation Mouse Development at the Single-Cell Level. *Cell,* 175, 859-876.e33.

Milletari, F., N. Navab & S.-A. Ahmadi. 2016. V-Net: Fully Convolutional Neural Networks for Volumetric Medical Image Segmentation. In *2016 Fourth International Conference on 3D Vision (3DV)*. IEEE.

Murray, J. I., Z. Bao, T. J. Boyle, M. E. Boeck, B. L. Mericle, T. J. Nicholas, Z. Zhao, M. J. Sandel & R. H. Waterston (2008) Automated analysis of embryonic gene expression with cellular resolution in C. elegans. *Nature Methods,* 5, 703-709.

Panaretos, V. M. & Y. Zemel (2019) Statistical Aspects of Wasserstein Distances. *Annual Review of Statistics and Its Application,* 6, 405-431.

Paszke, A., S. Gross, F. Massa, A. Lerer, J. Bradbury, G. Chanan, T. Killeen, Z. Lin, N. Gimelshein, A. D. Luca Antiga, E. Y. Andreas Kopf, Z. DeVito, M. Raison, A. Tejani, S. Chilamkurthy, B. Steiner, L. Fang, J. Bai & S. Chintala. 2019. PyTorch: An Imperative Style, High-Performance Deep Learning Library. In *Advances in Neural Information Processing Systems (NeurIPS 2019)*.

Ronneberger, O., P. Fischer, T. Brox, N. Navab, J. Hornegger, W. M. Wells & A. F. Frangi. 2015a. U-Net: Convolutional Networks for Biomedical Image Segmentation. In *Medical Image Computing and Computer-Assisted Intervention – MICCAI 2015*, 234-241. Springer International Publishing.

---. 2015b. U-Net: Convolutional Networks for Biomedical Image Segmentation. In *Medical Image Computing and Computer-Assisted Intervention – MICCAI 2015*. Springer International Publishing.
Schmidt, U., M. Weigert, C. Broaddus & G. Myers. 2018. Cell Detection with Star-Convex Polygons. In *Medical Image Computing and Computer Assisted Intervention – MICCAI 2018*, 265-273. Springer International Publishing.
Stegmaier, J., F. Amat, C. Lemon, William, K. Mcdole, Y. Wan, G. Teodoro, R. Mikut & J. Keller, Philipp (2016) Real-Time Three-Dimensional Cell Segmentation in Large-Scale Microscopy Data of Developing Embryos. *Developmental Cell,* 36**,** 225-240.
Stringer, C., T. Wang, M. Michaelos & M. Pachitariu (2021) Cellpose: a generalist algorithm for cellular segmentation. *Nature Methods,* 18**,** 100-106.
Sugawara, K., Ç. Çevrim & M. Averof (2022) Tracking cell lineages in 3D by incremental deep learning. *eLife,* 11.
Sugioka, K. & B. Bowerman (2018) Combinatorial Contact Cues Specify Cell Division Orientation by Directing Cortical Myosin Flows. *Developmental Cell,* 46**,** 257-270.e5.
Sulston, J. E., E. Schierenberg, J. G. White & J. N. Thomson (1983) The embryonic cell lineage of the nematode *Caenorhabditis elegans*. *Developmental biology,* 100**,** 64-119.
Van Der Walt, S., J. L. Schönberger, J. Nunez-Iglesias, F. Boulogne, J. D. Warner, N. Yager, E. Gouillart & T. Yu (2014) scikit-image: image processing in Python. *PeerJ,* 2**,** e453.
Vaswani, A., N. Shazeer, N. Parmar, J. Uszkoreit, L. Jones, A. N. Gomez, Ł. Kaiser & I. Polosukhin (2017) Attention is all you need. *Advances in neural information processing systems,* 30.
Waibel, D. J. E., S. Atwell, M. Meier, C. Marr & B. Rieck. 2022. Capturing Shape Information with Multi-scale Topological Loss Terms for 3D Reconstruction. In *Lecture Notes in Computer Science*, 150-159. Springer Nature Switzerland.
Wang, A., Q. Zhang, Y. Han, S. Megason, S. Hormoz, K. R. Mosaliganti, J. C. K. Lam & V. O. K. Li (2022) A novel deep learning-based 3D cell segmentation framework for future image-based disease detection. *Scientific Reports,* 12.
Weigert, M., U. Schmidt, R. Haase, K. Sugawara & G. Myers. 2020. Star-convex polyhedra for 3D object detection and segmentation in microscopy. In *Proceedings of the IEEE/CVF winter conference on applications of computer vision*, 3666-3673.
Zhou, Z., M. M. R. Siddiquee, N. Tajbakhsh & J. Liang (2020) UNet++: Redesigning Skip Connections to Exploit Multiscale Features in Image Segmentation. *IEEE Transactions on Medical Imaging,* 39**,** 1856-1867.

## Supplementary Figures

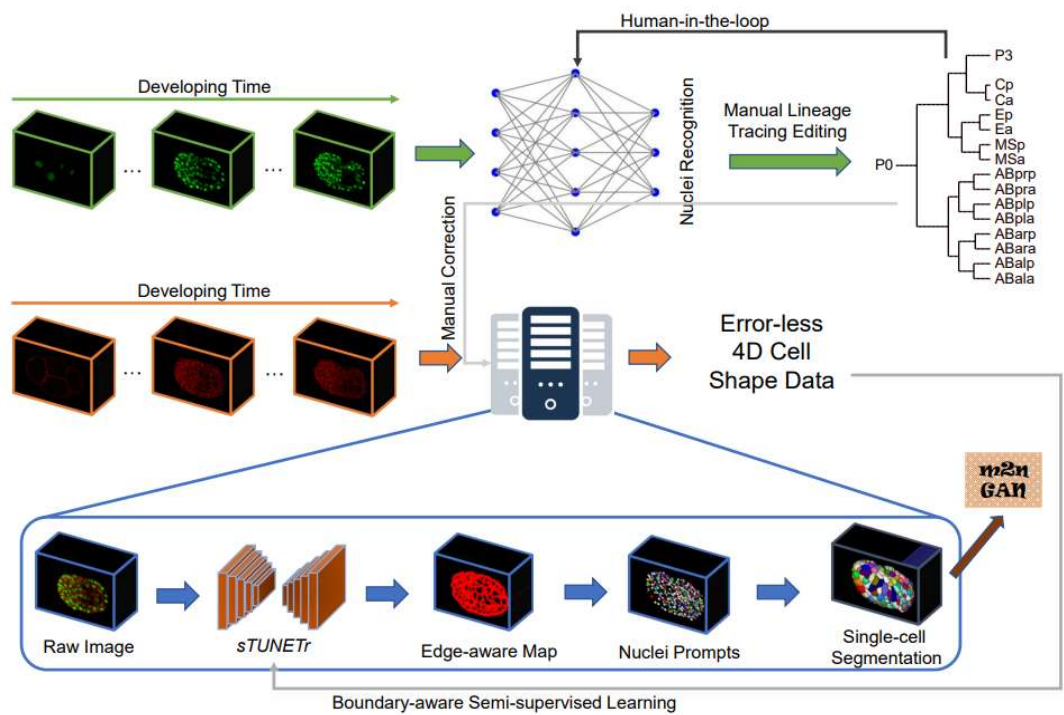

**Figure S1. The framework of *CTransformer* for complete cell morphology map to 550−cell stage with nuclei images assisted.** The human-in-the-loop workflow for semi-supervised learning strategy in training the single-cell segmentation process.

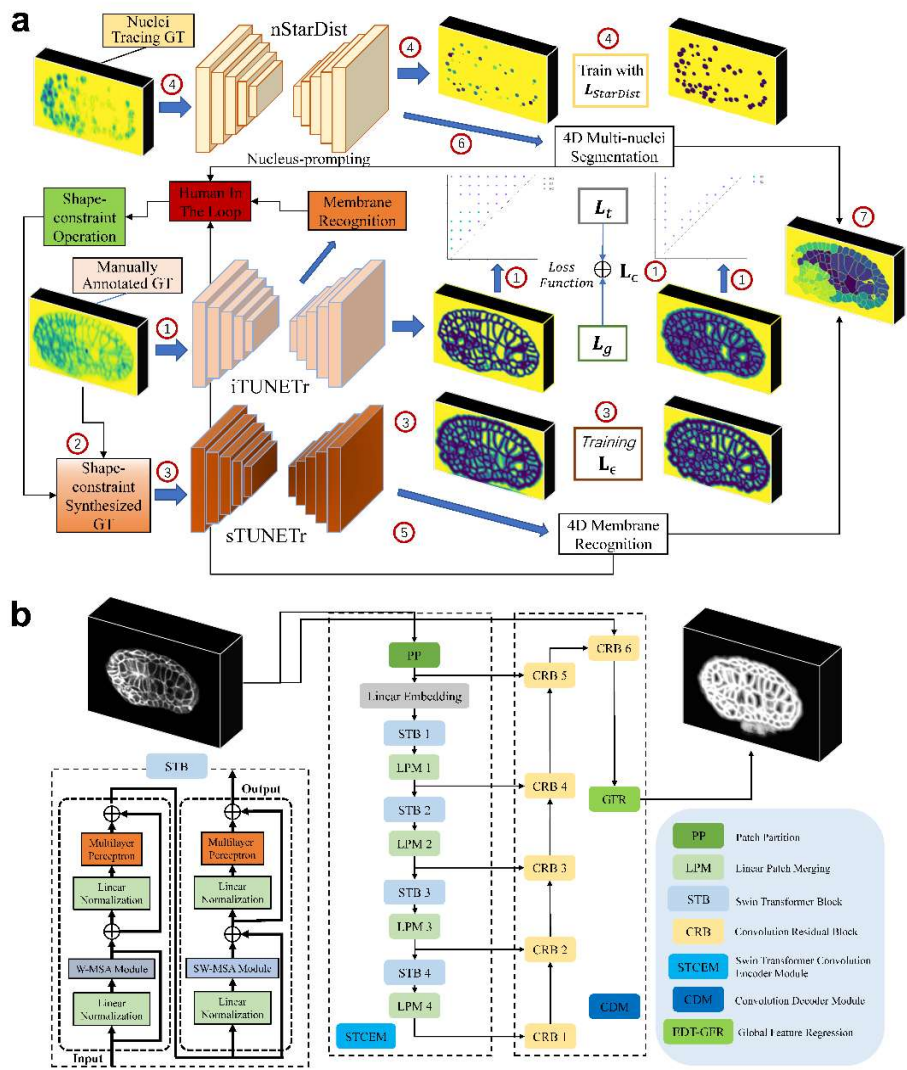

**Figure S2. The training process for the integrated DNN model, *TUNETr* of *CTransformer* framework. (a)** TUNETr of *CTransformer* is a well−designed DNN model for cell membrane recognition and multi−cell segmentation. In (**a**), from 1 to 3, and 5 steps, *CTransformer* trained 2 *TUNETr* models, *iTUNETr* and *sTUNETr* for 3D cell membrane recognition, and an adaptive StarDist3D model for 4D nuclei segmentation. In 4, 6 to 7 steps, *CTransformer* utilized the results of cellular membrane recognition, nuclei segmentation and Delaunay−watershed algorithm to produce the multiple whole−cell segmentation for 4D fluorescence images. (**b**) The figure illustrates the transformer, based on encoder−decoder DNN structure used in *iTUNETr* and *sTUNETr*. The most important part is the STB module, deploying attention mechanism in the DNN. The details of modules are explained in the Section Materials and Methods.

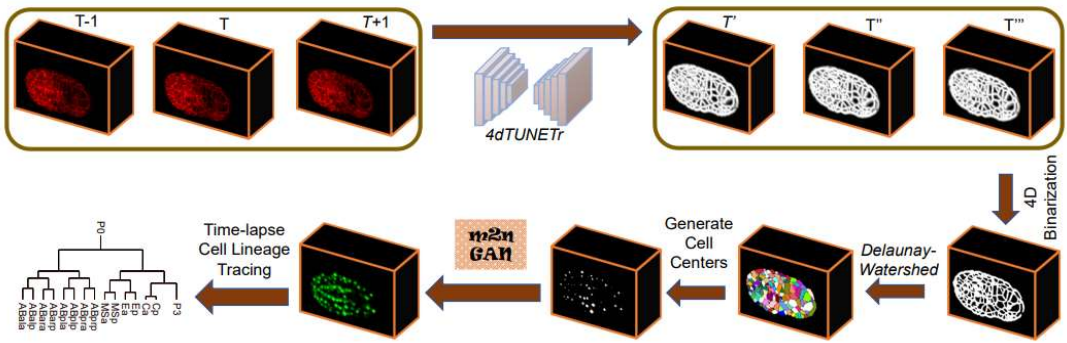

**Figure S3. The *4dTUNETr* and *m2nGAN* detailed running procedure.** *4dTUNETr* accepts 3 sequential raw membrane volumes as input and predicts the redundant membranes for one time point. Without cell shape centroids prompting, the *m2nGAN* generates pseudo nuclei images as the resource in lineage tracing.

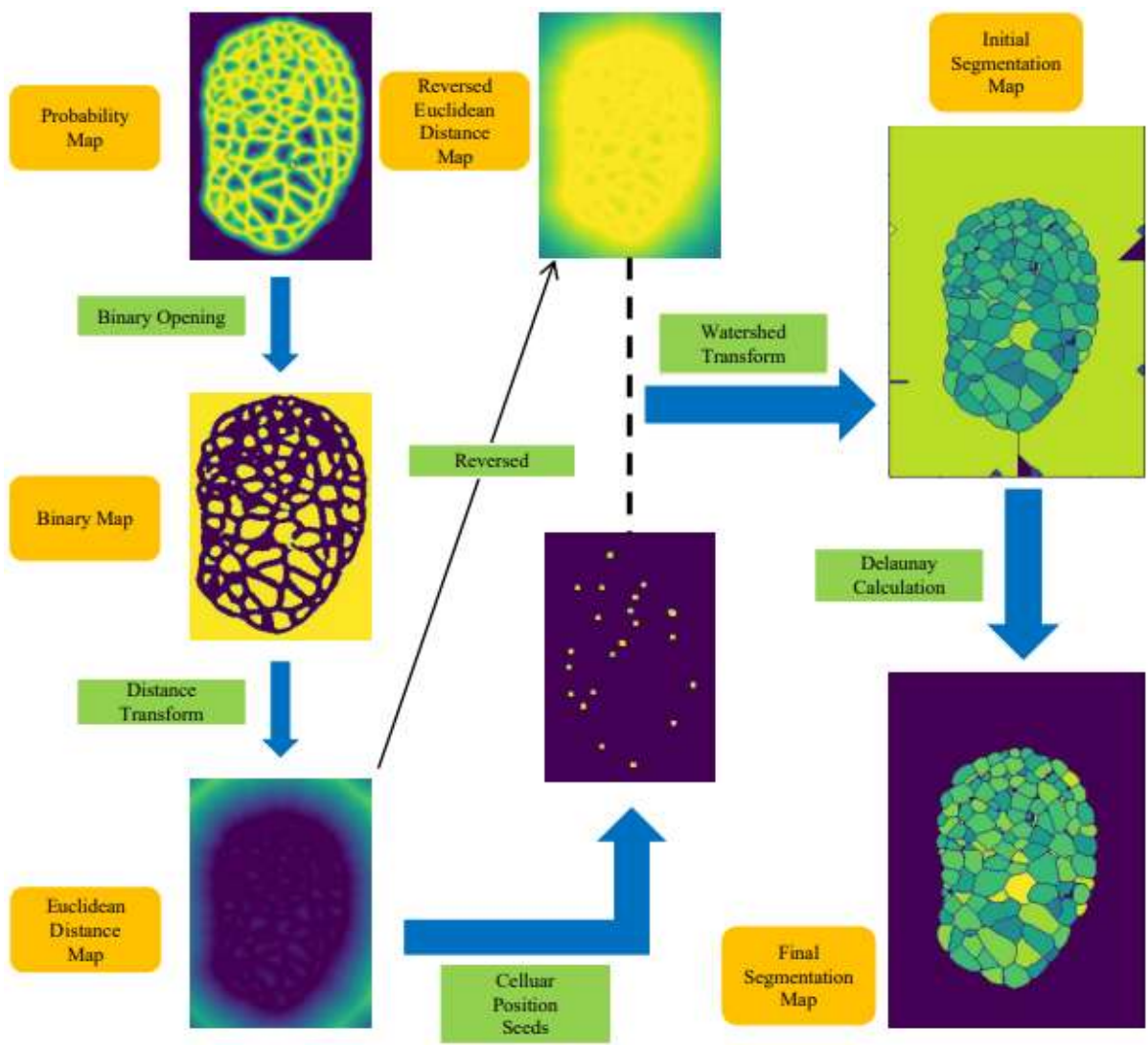

**Figure S4: The complete 3D images processing dataflow of running output *TUNETr*, Delaunay−Watershed graphical algorithm without nuclei prompting**. The dataflow is illustrated with 2D cut of the volumes. Cell membrane recognition (probability map prediction), binarized membrane segmentation, Euclidean distance transformed map, cellular position seeding map, multiple whole−cell segmentation, and Delaunay−processed segmentation are shown.

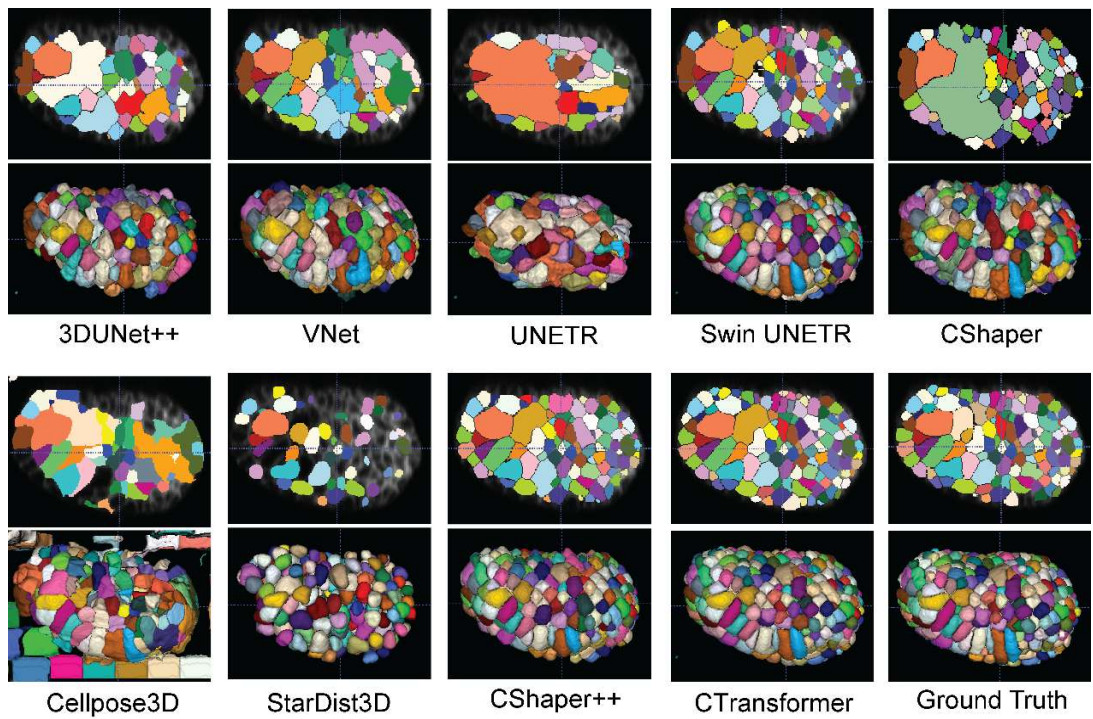

**Figure S5: Qualitative illustration between *TUNETr* and other existing SOTA segmentation algorithms**. 2D YZ view (upper row) and 3D side view (lower row) of the ground truth segmentation and segmentations generated from different DNN models, like VNet, UNETR, Cellpose3D, StarDist3D at 550−cell stage in *C. elegans*. Our results are generated by *sTUNETr*.

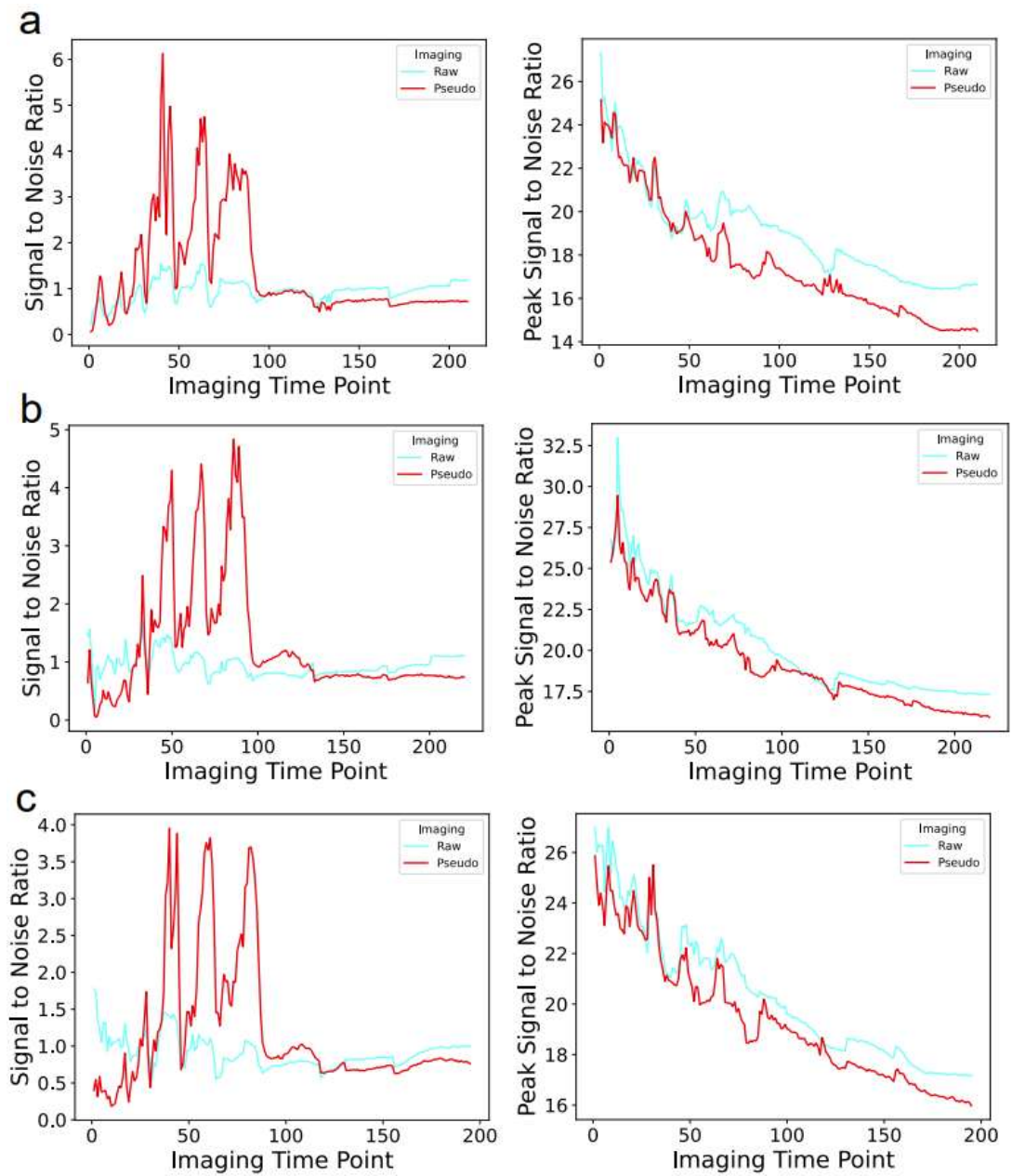

**Figure S6: Image-wise evaluation of another two embryos between pseudo and raw nuclei images.** Comparison of Signal to Noise Ratio (SNR) and Peak Signal to Noise (PSNR) between raw (cyan) and pseudo images (red) for one embryo across time. Shown are the values of the living embryo from 1 to 220 time point. Note that a higher SNR or PSNR depicts less signal discrepancies from ground truth images, and more signals and less noises for nuclei tracing. (**a**) The comparison of WT_C_Sample6. (**b**) The evaluation of WT_Sample4. (**c**) The results from WT_Sample5.

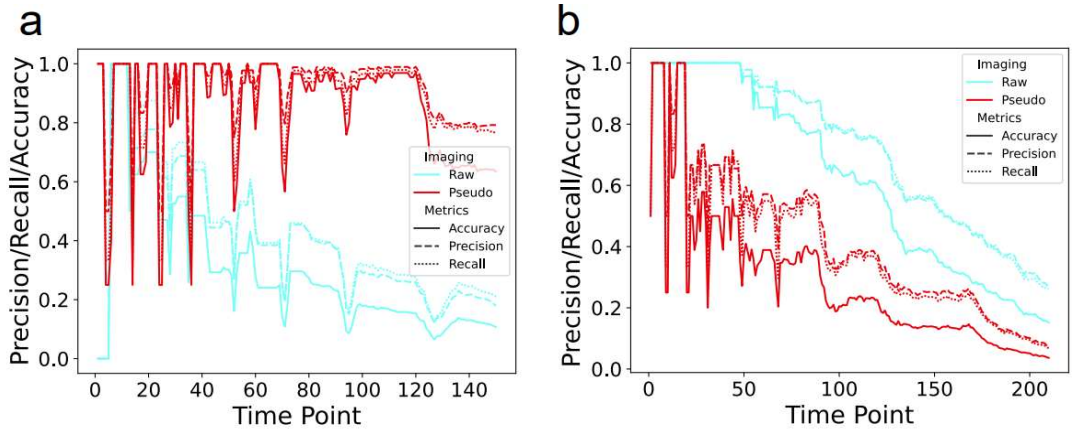

**Figure S7: Lineage tracing comparison, accuracy, precision and recall between generative and raw nuclei images on embryos (a) WT_Sample4 and (b) WT_C_Sample6, respectively, with StarryNite.**

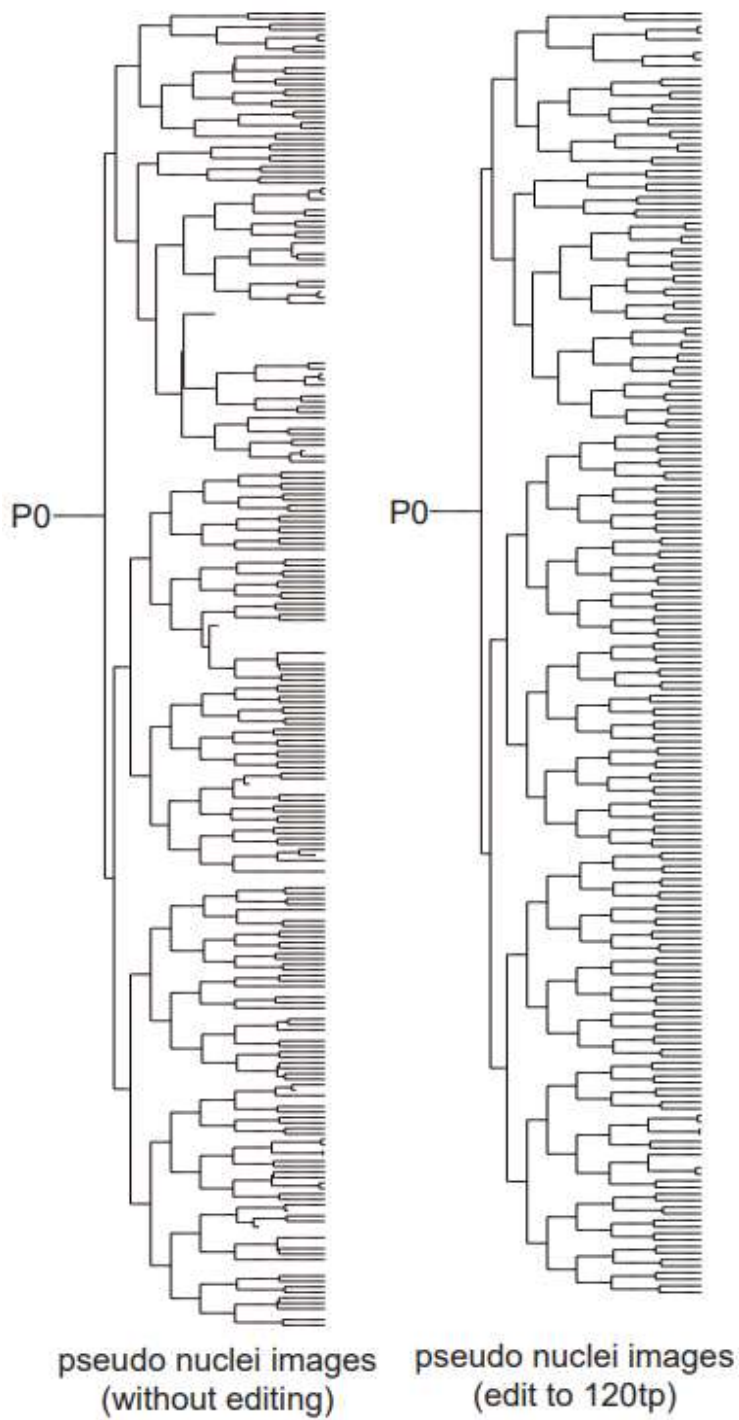

**Figure S8: Before and after edited on generative pseudo nuclei images generated on WT_Sample4 with StarryNite and edited by AceTree.** The edited cell lineage tree showcases the accuracy and facility of the produce linage tree with moderate number of tracking and tracing errors.

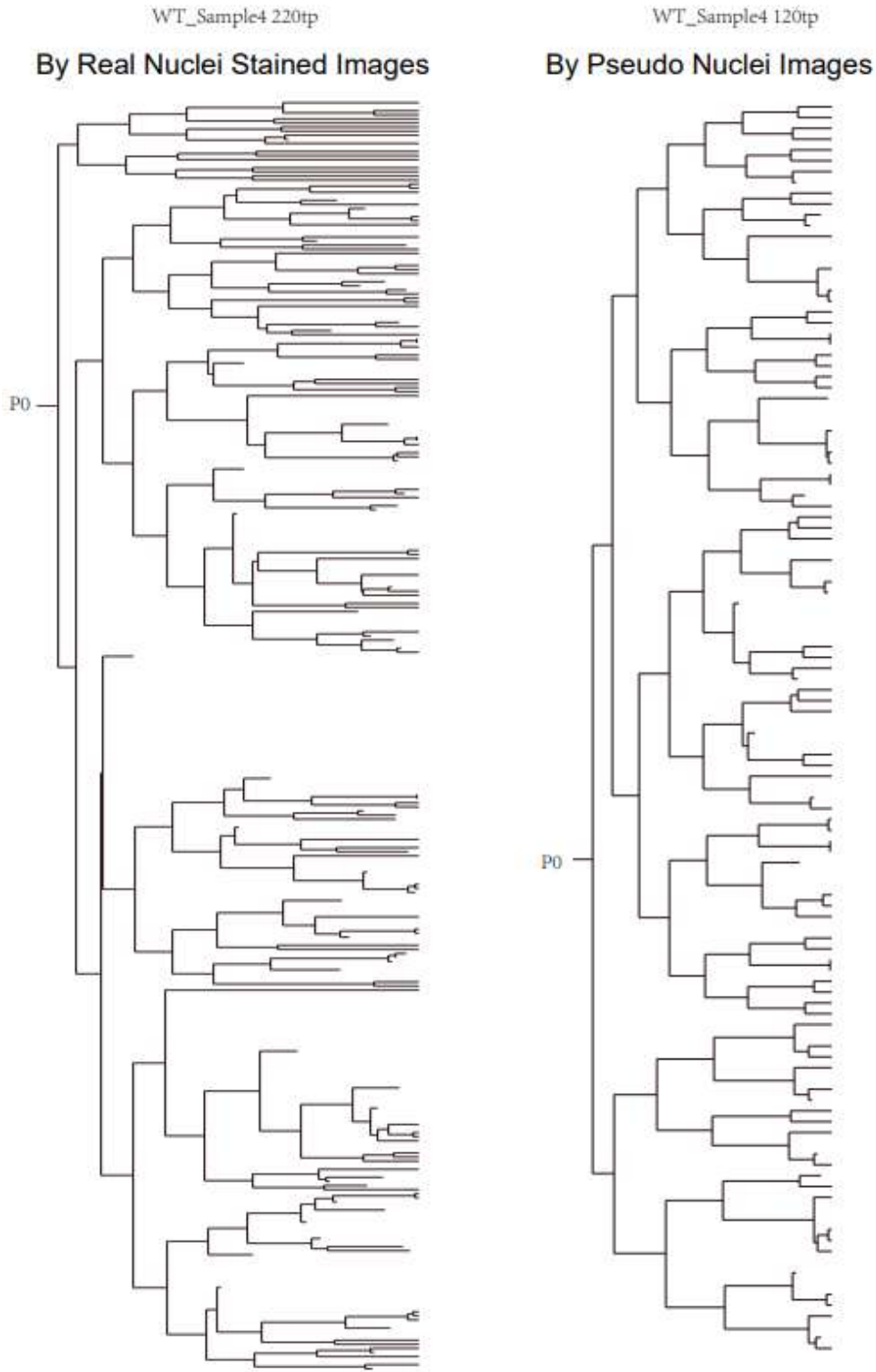

**Figure S9: Lineage−wise (the tree) comparison between generative and raw nuclei images on uncompressed embryo WT_Sample4 with StarryNite and AceTree.** P0 is the root and demonstrated to 120 time point without manual curation.

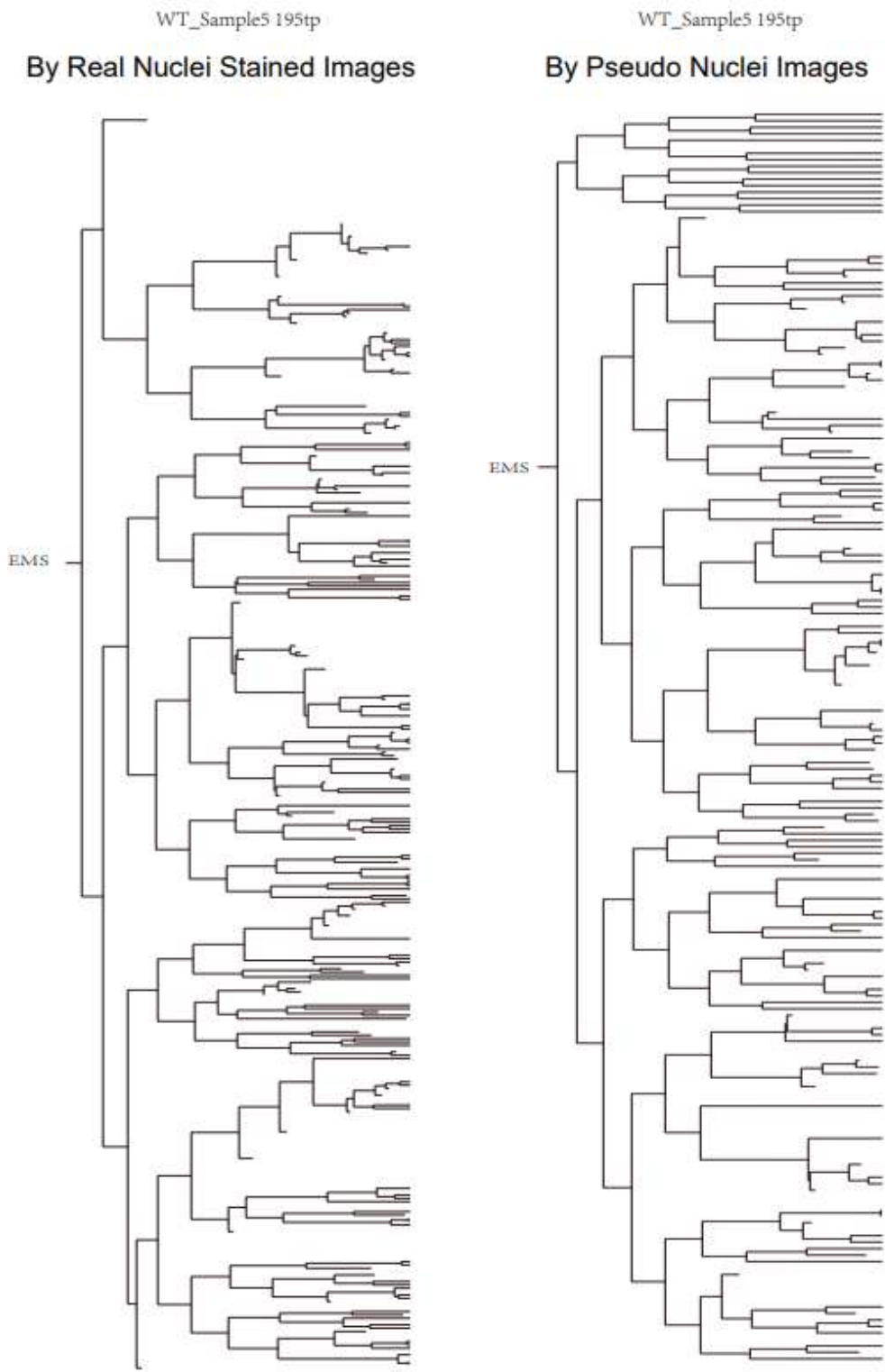

**Figure S10: Lineage−wise (the tree) comparison between generative and raw nuclei images on uncompressed embryo WT_Sample5 with StarryNite and AceTree.** EMS is the root and pushed to 195 time point, up to 550-cell stage without manual curation.

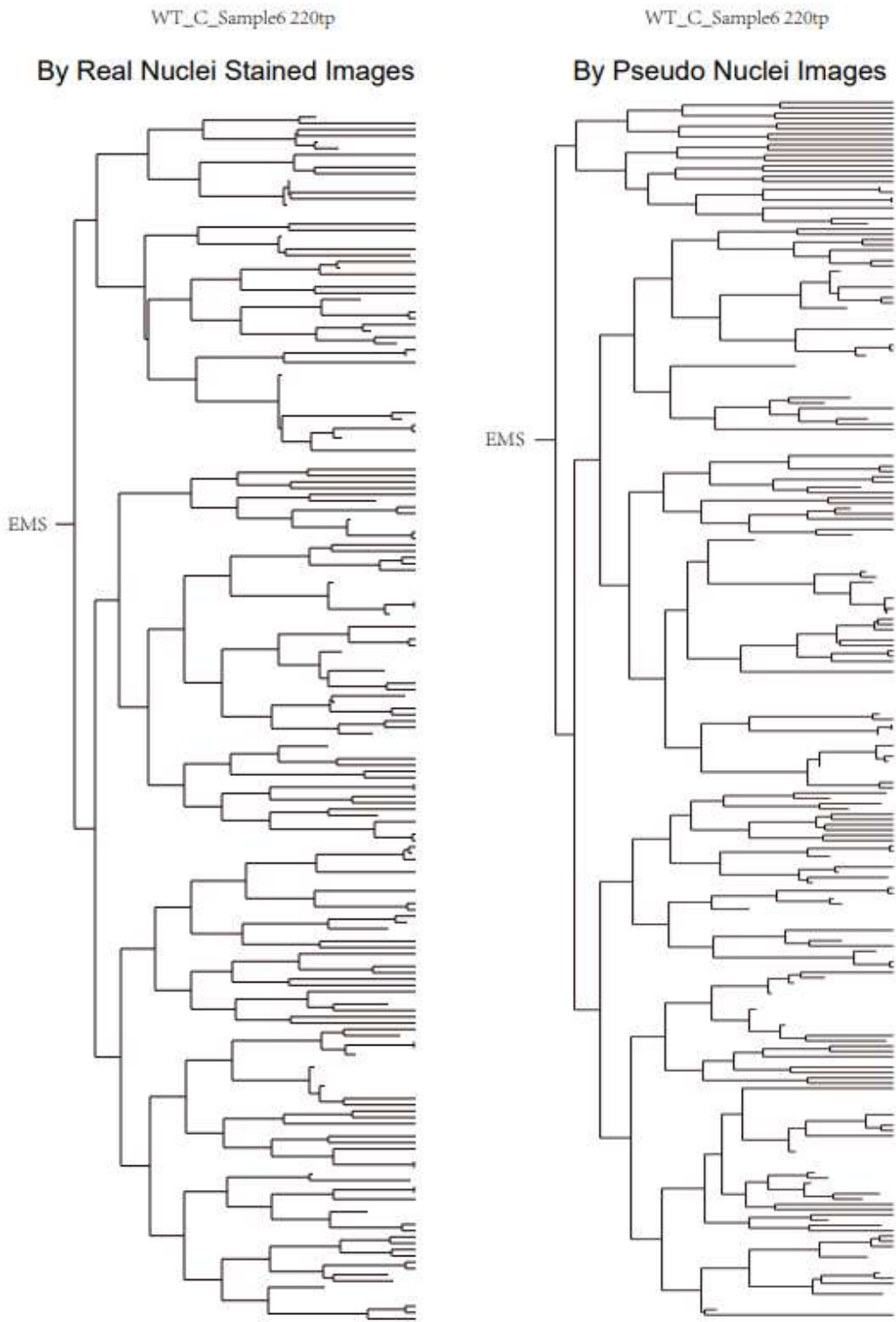

**Figure S11: Lineage−wise (the tree) comparison between generative and raw nuclei images on compressed embryo WT_C_Sample6 with StarryNite and AceTree.** EMS is the root and pushed to 210 time point, up to 530-cell stage without manual curation.

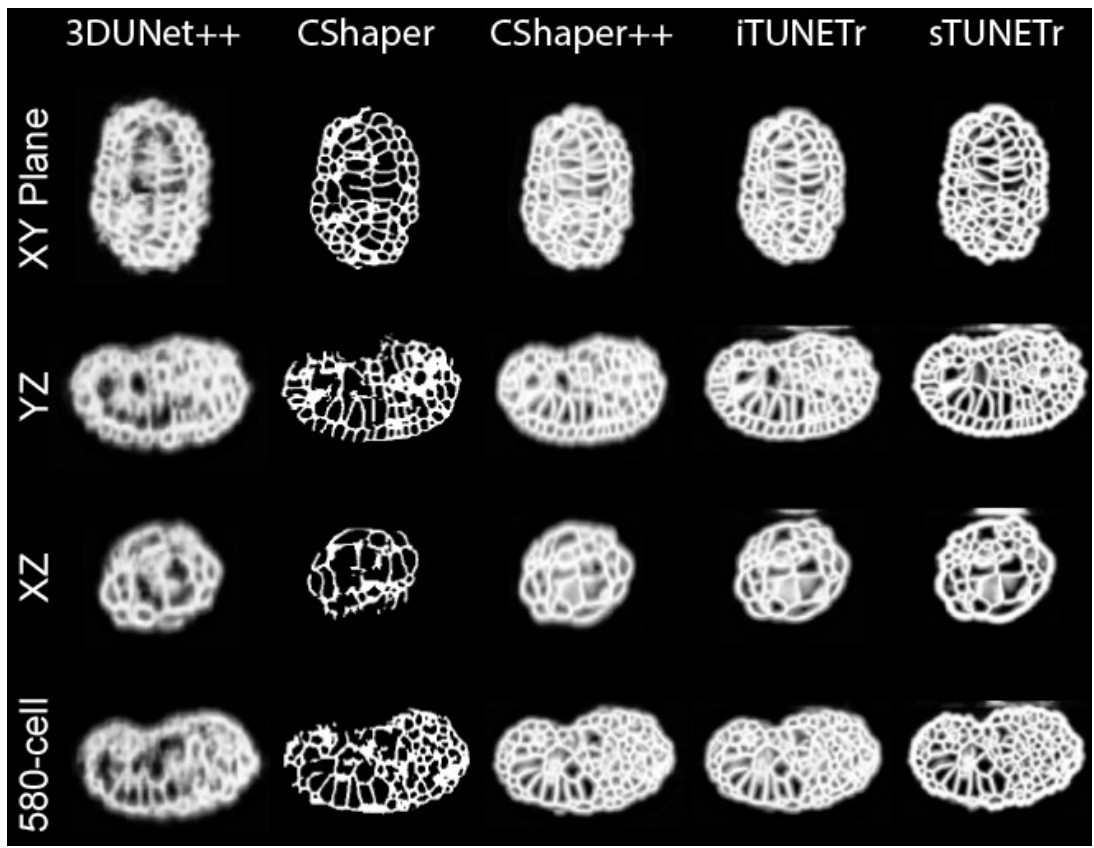

**Figure S12. Qualitative demonstrations membrane recognitions at 550- and 580-cell stage.** At XY (lateral) plane, first row shows the low membrane brightness because of intrinsic fluorescence heterogeneity. Axially, second row presents the significant loss of signal at the edge of the volume. Third row demonstrates the weak signals at bottom and weakened intensities along the laser direction. Fourth row, i.e. 580-cell stage embryo, illustrates severely fuzzy signals and undersampling from isotropic imaging. The five columns behind were giving membrane recognition results from 5 typical DNN models.

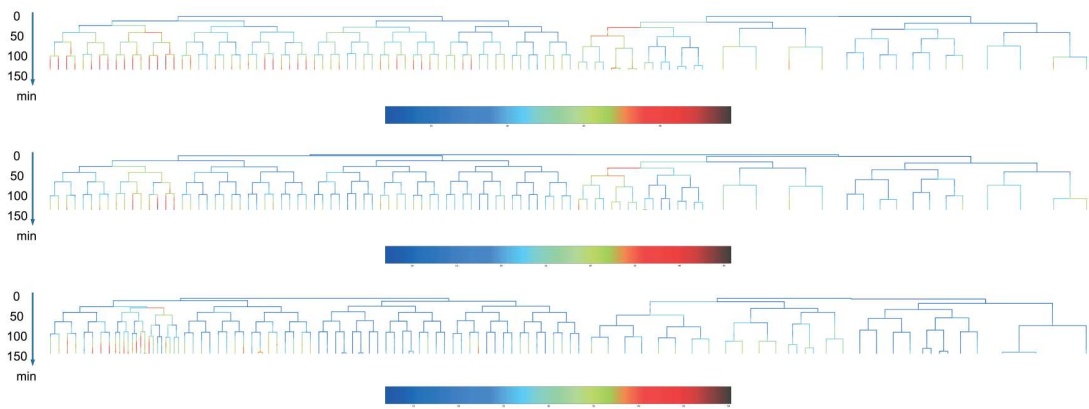

**Figure S13. The cell-wise adhesion demonstrations of three time-lapse embryos (quantitative analyses on *hmr-1*).**

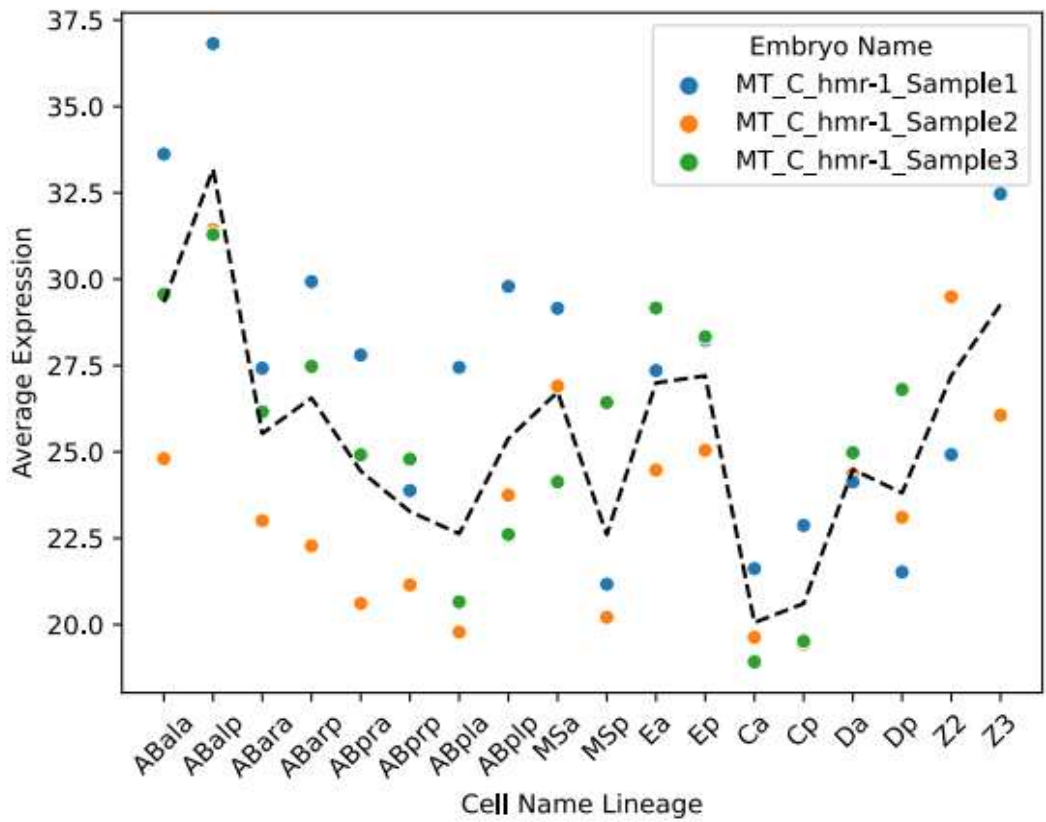

**Figure S14. Surface unit cell adhesion on 18 sublineages.**

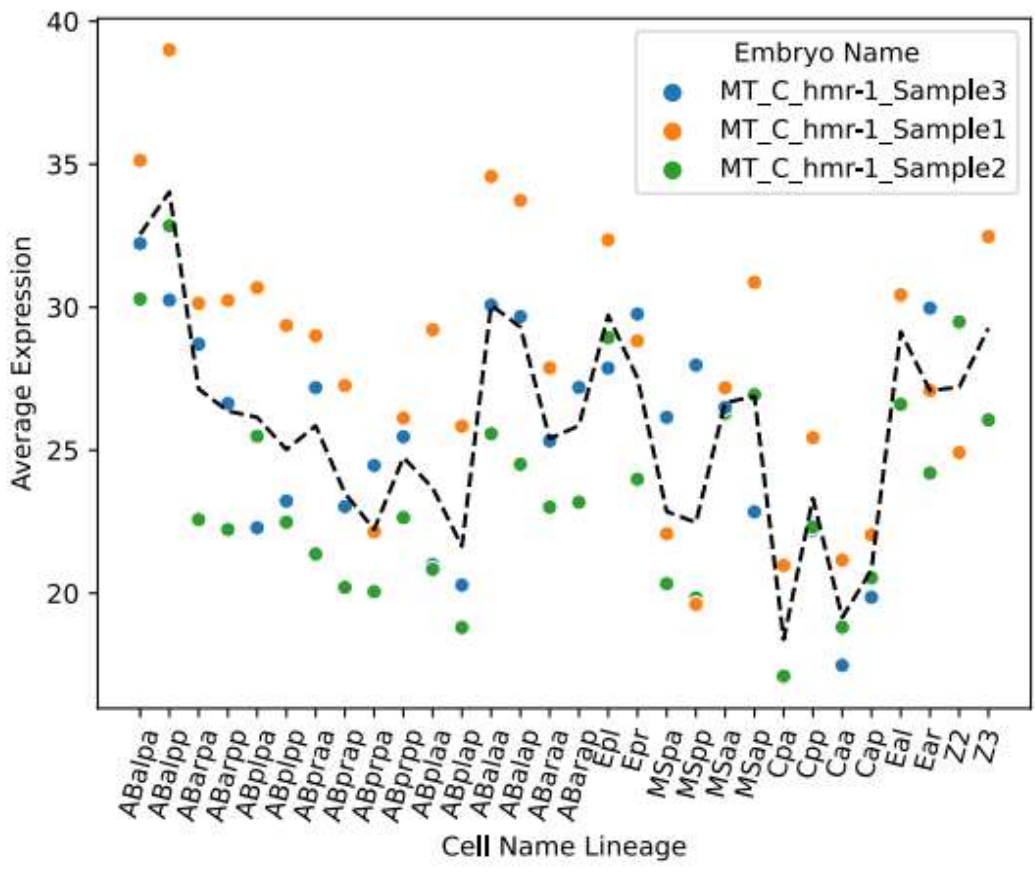

**Figure S15. Surface unit cell adhesion on 30 sublineages.**